\documentclass[aip,jcp,reprint,noshowkeys]{revtex4-1}
\usepackage{graphicx,dcolumn,bm,xcolor,microtype,multirow,amscd,amsmath,amssymb,amsfonts,physics,longtable,soul}
\usepackage[version=4]{mhchem}

\usepackage[normalem]{ulem}
\usepackage{soul}
\usepackage{mathpazo,libertine}
\usepackage[normalem]{ulem}

\definecolor{darkgreen}{RGB}{0, 180, 0}

\usepackage{hyperref}
\hypersetup{
    colorlinks=true,
    linkcolor=blue,
    filecolor=blue,      
    urlcolor=blue,	
    citecolor=blue
}

\newcommand{\cdash}{\multicolumn{1}{c}{---}}
\newcommand{\mc}{\multicolumn}

\newcommand{\mcc}[1]{\multicolumn{1}{c}{#1}}

\newcommand{\SI}{\textcolor{blue}{supporting information}}
	
\newcommand{\EPT}{E_\text{PT2}}
\newcommand{\EFCI}{E_\text{FCI}}
\newcommand{\EsCI}{E_\text{sCI}}

\newcommand{\ex}[4]{$^{#1}#2_{#3}^{#4}$}
\renewcommand{\tr}[2]{$(#1)\rightarrow(#2)$}

\newcommand{\IneV}[1]{#1~eV}
\newcommand{\InAU}[1]{#1~a.u.}

\newcommand{\ra}{\rightarrow}
\newcommand{\pis}{\pi^\star}
\newcommand{\si}{\sigma}
\newcommand{\sis}{\sigma^\star}
\newcommand{\nO}{n_\text{O}}
\newcommand{\nN}{n_\text{N}}
\newcommand{\piCC}{\pi_{\ce{CC}}}
\newcommand{\pisCC}{\pi_{\ce{CC}}^\star}
\newcommand{\piCO}{\pi_{\ce{CO}}}
\newcommand{\pisCO}{\pi_{\ce{CO}}^\star}
\newcommand{\siCC}{\si_{\ce{CC}}}
\newcommand{\sisCC}{\si_{\ce{CC}}^\star}
\newcommand{\siCO}{\si_{\ce{CO}}}
\newcommand{\sisCO}{\si_{\ce{CO}}^\star}

\newcommand{\LCPQ}{Laboratoire de Chimie et Physique Quantiques (UMR 5626), Universit\'e de Toulouse, CNRS, UPS, France}
\newcommand{\CEISAM}{Laboratoire CEISAM (UMR 6230), CNRS, Universit\'e de Nantes, Nantes, France}

\begin{document}	

\title{Reference Energies for Double Excitations}

\author{Pierre-Fran\c{c}ois Loos}
\email[Corresponding author: ]{loos@irsamc.ups-tlse.fr}
\affiliation{\LCPQ}
\author{Martial Boggio-Pasqua}
\affiliation{\LCPQ}
\author{Anthony Scemama}
\affiliation{\LCPQ}
\author{Michel Caffarel}
\affiliation{\LCPQ}
\author{Denis Jacquemin}
\affiliation{\CEISAM}    

\begin{abstract}
Excited states exhibiting double excitation character are notoriously difficult to model using conventional single-reference methods, such as adiabatic time-dependent density-functional theory (TD-DFT) or equation-of-motion coupled cluster (EOM-CC).  
In addition, these states are typical experimentally ``dark'' making their detection in photo-absorption spectra very challenging.  
Nonetheless, they play a key role in the faithful description of many physical, chemical, and biological processes.  
In the present work, we provide accurate reference excitation energies for transitions involving a substantial amount of double excitation using a series of increasingly large diffuse-containing atomic basis sets.
Our set gathers 20 vertical transitions from 14 small- and medium-size molecules (acrolein, benzene, beryllium atom, butadiene, carbon dimer and trimer, ethylene, formaldehyde, glyoxal, hexatriene, nitrosomethane, nitroxyl, pyrazine, and tetrazine).  
Depending on the size of the molecule, selected configuration interaction (sCI) and/or multiconfigurational (CASSCF, CASPT2, (X)MS-CASPT2 and NEVPT2) calculations are performed in order to obtain reliable estimates of the vertical transition energies.  
In addition, coupled cluster approaches including at least contributions from iterative triples (such as CC3, CCSDT, CCSDTQ, and CCSDTQP) are assessed.  
Our results clearly evidence that the error in CC methods is intimately related to the amount of double excitation character of the transition.  
For ``pure'' double excitations (i.e.~for transitions which do not mix with single excitations), the error in CC3 can easily reach \IneV{$1$}, while it goes down to few tenths of an eV for more common transitions (like in \emph{trans}-butadiene) involving a significant amount of singles.  
As expected, CC  approaches including quadruples yield highly accurate results for any type of transitions.  
The quality of the excitation energies obtained with multiconfigurational methods is harder to predict.  
We have found that the overall accuracy of these methods is highly dependent of both the system and the selected active space. 
The inclusion of the $\si$ and $\sis$ orbitals in the active space, even for transitions involving mostly $\pi$ and $\pis$ orbitals, is mandatory in order to reach high accuracy. 
A theoretical best estimate (TBE) is reported for each transition. 
We believe that these reference data will be valuable for future methodological developments aiming at accurately describing double excitations.
\\
\begin{center}
	\includegraphics[width=0.4\linewidth]{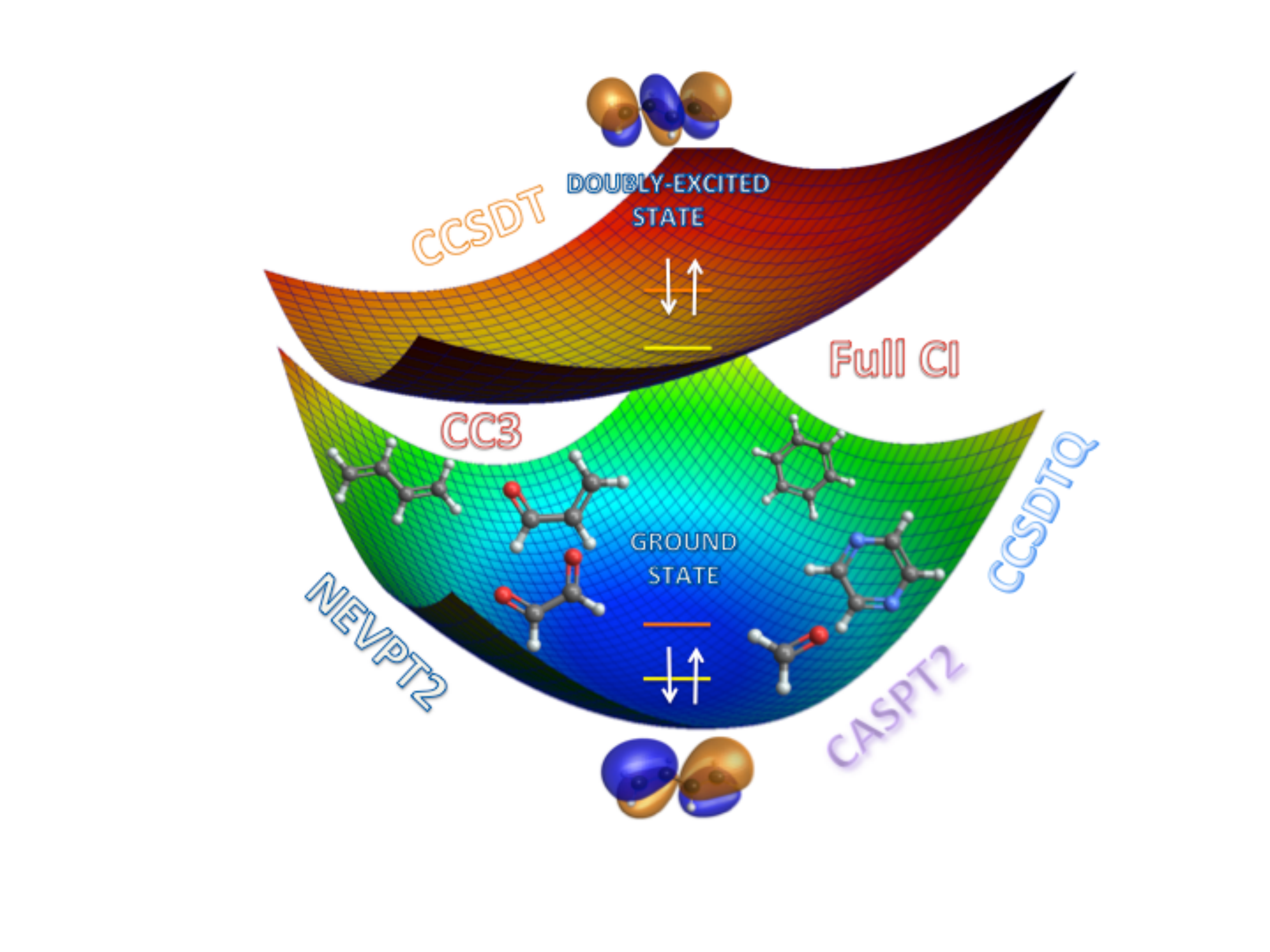}
	\\
	\bf TOC graphical abstract
\end{center}
\end{abstract}

\maketitle

%%%%%%%%%%%%%%%%%%%%%%%%
\section{
Introduction
\label{sec:intro}
}
%%%%%%%%%%%%%%%%%%%%%%%%

Within the theoretical and computational quantum chemistry community, the term \textit{``double excitation''} commonly refers to a state whose configuration interaction (CI) or coupled cluster (CC) expansion includes \textit{significant} coefficients or amplitudes associated to doubly-excited Slater determinants, i.e., determinants in which two electrons have been promoted from occupied to virtual orbitals of the chosen \textit{reference} determinant.   
Obviously, this definition is fairly ambiguous as it is highly dependent on the actual reference Slater determinant, and on the magnitude associated with the term ``\textit{significant}''.  
Moreover, such a picture of placing electrons in orbitals only really applies to one-electron theories, e.g., Hartree-Fock \cite{SzaboBook} or Kohn-Sham. \cite{ParrBook} 
In contrast, in a many-electron picture, an excited state is a linear combination of Slater determinants usually built from an intricate mixture of single, double and higher excitations.   
In other words, the definition of a double excitation remains fuzzy, and this has led to controversies regarding the nature of the 2\,\ex{1}{A}{1g}{} and 1\,\ex{1}{E}{2g}{} excited states of butadiene \cite{Shu_2017,Barca_2018a,Barca_2018b} and benzene, \cite{Barca_2018a,Barca_2018b} respectively, to mention two well-known examples. 
Although these two states have been classified as doubly-excited states in the past, Barca et al.~have argued that they can be seen as singly-excited states if one allows sufficient orbital relaxation in the excited state. \cite{Barca_2018a,Barca_2018b} 
Nonetheless, in the remaining of this paper, we will follow one of the common definition and define a double excitation as an excited state with a significant amount of double excitation character in the multideterminant expansion.

Double excitations do play a significant role in the proper description of several key physical, chemical and biological processes, e.g., in photovoltaic devices, \cite{Delgado_2010} in the photophysics of vision, \cite{Palczewski_2006} and in photochemistry in general \cite{Bernardi_1990,Bernardi_1996,Boggio-Pasqua_2007,Klessinger_1995,Olivucci_2010,Robb_2007,VanderLugt_1969} involving ubiquitous conical intersections. \cite{Levine_2006}
The second example is intimately linked to the correct location of the excited states of polyenes, \cite{Serrano-Andres_1993,Cave_1988b,Lappe_2000,Boggio-Pasqua_2004,Maitra_2004,Cave_2004,Wanko_2005,Starcke_2006,Catalan_2006,Mazur_2009,Angeli_2010,Mazur_2011,Huix-Rotllant_2011} 
that are closely related to rhodopsin which is involved in visual phototransduction. \cite{Gozem_2012,Gozem_2013,Gozem_2013a,Gozem_2014,Huix-Rotllant_2010,Xu_2013,Schapiro_2014,Tuna_2015,Manathunga_2016}   
Though doubly-excited states do not appear directly in photo-absorption spectra, these dark states strongly mix with the bright singly-excited states leading to the formation of satellite peaks. \cite{Helbig_2011,Elliott_2011}

From a theoretical point of view, double excitations are notoriously difficult to model using conventional single-reference methods. \cite{Sundstrom_2014}
For example, the adiabatic approximation of time-dependent density-functional theory (TD-DFT) \cite{Casida} yields reliable excitation spectra with great efficiency in many cases.
Nevertheless, fundamental deficiencies in TD-DFT have been reported for the computation of extended conjugated systems, \cite{Woodcock_2002,Tozer_2003} charge-transfer states, \cite{Tozer_1999,Dreuw_2003,Sobolewski_2003,Dreuw_2004} Rydberg states,\cite{Tozer_1998,Tozer_2000,Casida_1998,Casida_2000,Tozer_2003} conical intersections, \cite{Tapavicza_2008,Levine_2006} and, more importantly here, for states with double excitation character. \cite{Levine_2006,Tozer_2000,Elliott_2011}
Although, using range-separated hybrids \cite{Tawada_2004,Yanai_2004} provides an effective solution to the first three cases, one must go beyond the ubiquitous adiabatic approximation to capture the latter two.
One possible solution is provided by spin-flip TD-DFT which describes double excitations as single excitations from the lowest triplet state. \cite{Huix-Rotllant_2010,Krylov_2001,Shao_2003,Wang_2004,Wang_2006,Minezawa_2009}
However, major limitations pertain. \cite{Huix-Rotllant_2010} In order to go beyond the adiabatic approximation, a dressed TD-DFT approach has been proposed by Maitra and coworkers \cite{Maitra_2004,Cave_2004} (see also Refs.~\onlinecite{Mazur_2009,Mazur_2011,Huix-Rotllant_2011,Elliott_2011,Maitra_2012}).
In this approach the exchange-correlation kernel is made frequency dependent, \cite{Romaniello_2009a,Sangalli_2011} which allows to treat doubly-excited states.
Albeit far from being a mature black-box approach, ensemble DFT \cite{Theophilou_1979,Gross_1988,Gross_1988a,Oliveira_1988} is another viable alternative currently under active development. \cite{Kazaryan_2008,Filatov_2015,Senjean_2015,Filatov_2015b,Filatov_2015c,Deur_2017,Gould_2018,Sagredo_2018} 

As shown by Watson and Chan, \cite{Watson_2012} one can also rely on high-level truncation of the equation-of-motion (EOM) formalism of CC theory in order to capture double excitations. \cite{Hirata_2000,Sundstrom_2014} 
However, in order to provide a satisfactory level of correlation for a doubly-excited state, one must, at least, introduce contributions from the triple excitations in the CC expansion. 
In practice, this is often difficult as the scalings of CC3, \cite{Christiansen_1995b,Koch_1997} CCSDT, \cite{Noga_1987} and CCSDTQ \cite{Kucharski_1991} are $N^7$, $N^8$ and $N^{10}$, respectively (where $N$ is the number of basis functions), obviously limiting the applicability of 
this strategy to small molecules.

Multiconfigurational methods constitute a more natural class of methods to properly treat double excitations.
Amongst these approaches, one finds complete active space self-consistent field (CASSCF), \cite{Roos} its second-order perturbation-corrected variant (CASPT2), \cite{Andersson_1990} as well as the second-order $n$-electron valence state perturbation theory (NEVPT2). \cite{Angeli_2001a, Angeli_2001b, Angeli_2002} 
However, the exponential scaling of such methods with the number of active electrons and orbitals also limits their application to small active spaces in their traditional implementation, although using sCI as an active-space solver allows to target much larger active spaces. \cite{Smith_2017}

Alternatively to CC and multiconfigurational methods, one can also compute transition energies for both singly- and doubly-excited states using selected configuration interaction (sCI) methods 
\cite{Bender_1969,Whitten_1969,Huron_1973,Evangelisti_1983, Cimiraglia_1985, Cimiraglia_1987, Illas_1988, Povill_1992} which have recently demonstrated their ability to reach near full CI (FCI) quality energies for small molecules.
\cite{Abrams_2005,Bunge_2006,Bytautas_2009,Giner_2013,Caffarel_2014,Giner_2015,Garniron_2017b,Caffarel_2016,Holmes_2016,Sharma_2017,Holmes_2017,Chien_2018,Scemama_2018a,Scemama_2018b,Loos_2018,Garniron_2018,Evangelista_2014,Schriber_2016,Tubman_2016,Liu_2016,Per_2017,Ohtsuka_2017,Zimmerman_2017}
The idea behind such methods is to avoid the exponential increase of the size of the CI expansion by retaining the most energetically relevant determinants only, thanks to the use of a second-order energetic criterion to select perturbatively 
determinants in the FCI space. \cite{Giner_2013,Giner_2015,Caffarel_2016,Scemama_2018a,Scemama_2018b,Garniron_2018,Sharma_2017,Blunt_2018}

By systematically increasing the order of the CC expansion, the number of determinants in the sCI expansion as well as the size of the one-electron basis set, some of us have recently defined a reference series of more than 100 very 
accurate vertical transition energies in 18 small compounds. \cite{Loos_2018}  However, this set is constituted almost exclusively of single excitations. Here, we report accurate reference excitation energies for double excitations obtained 
with both sCI and multiconfigurational methods for a significant number of small- and medium-size molecules using various diffuse-containing basis sets. Moreover, the accuracy obtained with several coupled cluster approaches including, 
at least,  triple excitations are assessed.  We believe that these reference data are particularly valuable for future developments of methods aiming at accurately describing double excitations.

This manuscript is organized as follows.  Computational details are reported in Sec.~\ref{sec:comp} for EOM-CC (Sec.~\ref{sec:CC}), multiconfigurational (Sec.~\ref{sec:CAS}) and sCI (Sec.~\ref{sec:sCI}) methods.  
In Section \ref{sec:res}, we discuss our results for each compound and report a list of theoretical best estimates (TBEs) for each transition. 
We further discuss the overall performance of the different methods and draw our conclusions in Sec.~\ref{sec:ccl}.

%%% FIG 1 %%%
\begin{figure}
	\includegraphics[width=\linewidth]{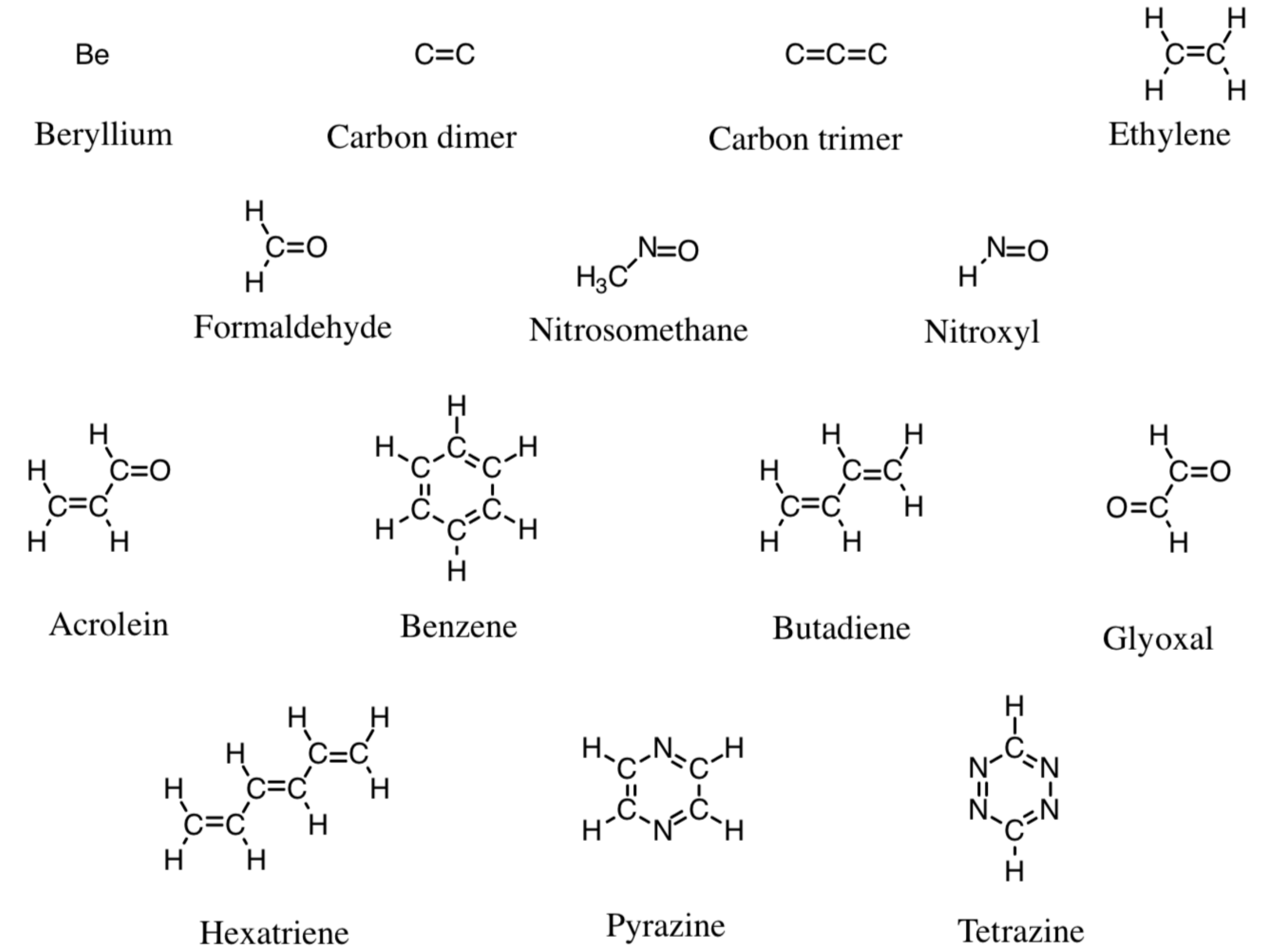}
	\caption{
	Structure of the various molecules considered in the present set. 
	\label{fig:mol}
	}
\end{figure}

%%%%%%%%%%%%%%%%%%%%%%%%
\section{
Computational details
\label{sec:comp}
}
%%%%%%%%%%%%%%%%%%%%%%%%

All geometries used in the present study are available in {\SI}.  They have been obtained at the CC3/aug-cc-pVTZ level (except for hexatriene where the geometry has been optimized at the CCSD(T)/aug-cc-pVTZ level) 
without applying the frozen core approximation following the same protocol as in earlier works where additional details can be found. \cite{Budzak_2017,Loos_2018} These geometry optimizations were performed with 
DALTON \cite{dalton} or CFOUR. \cite{cfour}  The so-called $\%T_1$ metric giving the percentage of single excitation calculated at the CC3 level in DALTON is employed to characterize the various states. For all calculations, we use 
the well-known Pople's 6-31+G(d) (with spherical gaussians) and Dunning's aug-cc-pVXZ (X $=$ D, T and Q) atomic basis sets.  In the following, we employ the AVXZ shorthand notations for Dunning's basis sets.

%---------------------------------------------
\subsection{
Coupled cluster calculations
\label{sec:CC}
}
%---------------------------------------------

Unless otherwise stated, the CC transition energies \cite{Kallay_2004} were computed in the frozen-core approximation.  
Globally, we used DALTON \cite{dalton} to perform the CC3 calculations, \cite{Christiansen_1995b,Koch_1997} CFOUR \cite{cfour} for the CCSDT \cite{Noga_1987} calculations, and MRCC \cite{mrcc} for CCSDT, \cite{Noga_1987} CCSDTQ, \cite{Kucharski_1991} (and higher) calculations. 
Because CFOUR and MRCC rely on different algorithms to locate excited states, we have interchangeably used these two softwares for the CCSDT calculations depending on the targeted transition.
Default program setting were generally applied, and when modified they have been tightened.  
Note that transition energies are identical in the EOM and linear response (LR) CC formalisms.  
Consequently, for the sake of brevity, we do not specify the EOM and LR terms in the remaining of this study. 
The total energies of all CC calculations are available in {\SI}.

%---------------------------------------------
\subsection{
Multiconfigurational calculations
\label{sec:CAS}
}
%---------------------------------------------

State-averaged (SA) CASSCF and CASPT2 \cite{Roos, Andersson_1990} have been performed with MOLPRO. \cite{molpro} 
Concerning the NEVPT2 calculations, the partially-contracted (PC) and strongly-contracted (SC) variants have been systematically tested. \cite{Angeli_2001a, Angeli_2001b, Angeli_2002} 
From a strict theoretical point of view, we point out that PC-NEVPT2 is supposed to be more accurate than SC-NEVPT2 given that it has a larger number of perturbers and greater flexibility.  
Additional information and technical details about the CASSCF (as well as CASSCF excitation energies), CASPT2 and NEVPT2 calculations can be found in {\SI}.
When there is a strong mixing between states with same spin and spatial symmetries, we have also performed calculations with multi-state (MS) CASPT2, \cite{Finley_1998} and its extended variant (XMS-CASPT2). \cite{Shiozaki_2011}
Unless otherwise stated, all CASPT2 calculations have been performed with level shift and IPEA parameters set to the standard values of $0.3$ and \InAU{$0.25$}, respectively. 

%---------------------------------------------
\subsection{
Selected configuration interaction calculations
\label{sec:sCI}
}
%---------------------------------------------

The sCI calculations reported here employ the CIPSI (Configuration Interaction using a Perturbative Selection made Iteratively) \cite{Huron_1973, Evangelisti_1983, Giner_2013} algorithm. 
We refer the interested reader to Refs.~\onlinecite{Caffarel_2016b, Scemama_2018a, Scemama_2018b, Garniron_2018, Loos_2018} for more details about sCI methods, and the CIPSI algorithm in particular.

In order to treat the electronic states on equal footing, a common set of determinants is selected for the ground state and excited states.  
These calculations can then be classified as ``state-averaged'' sCI. Moreover, to speed up convergence to the FCI limit, a common set of natural orbitals issued from a preliminary sCI calculation is employed. 
For the largest systems, few iterations might be required to obtain a well-behaved convergence of the excitation energies with respect to the number of determinants.  
For a given atomic basis set, we estimate the FCI limit by linearly extrapolating the sCI energy $\EsCI$ as a function of the second-order perturbative correction $\EPT$ which is an estimate of the truncation error in the sCI algorithm, i.e., $\EPT \approx \EFCI - \EsCI$. 
When $\EPT = 0$, the FCI limit has effectively been reached. 
To provide an estimate of the extrapolation error, we report the energy difference between the excitation energies obtained with two- and three-point linear fits. 
It is, however, a rough estimate as there is no univocal method to quantitatively measure the extrapolation error. 
This extrapolation procedure has nevertheless been shown to be robust, even for challenging chemical situations. \cite{Holmes_2017, Sharma_2017, Scemama_2018a, Scemama_2018b, Chien_2018, Garniron_2018, Loos_2018} 
In the following, these extrapolated sCI results are labeled exFCI. 
Here, $\EPT$ has been efficiently evaluated with a recently proposed hybrid stochastic-deterministic algorithm. \cite{Garniron_2017b}  
Note that we do not report error bars associated with $\EPT$ because the statistical errors originating from this algorithm are orders of magnitude smaller than the extrapolation errors.  

All the sCI calculations have been performed in the frozen core approximation with the electronic structure software QUANTUM PACKAGE, developed in Toulouse and freely available. \cite{QP}  
For the largest molecules considered here, our sCI wave functions contain up to $2 \times 10^8$ determinants which corresponds to an increase of two orders of magnitude compared to our previous study. \cite{Loos_2018}
Additional information about the sCI wave functions, excitations energies as well as their extrapolated values can be found in {\SI}.

%%% TABLE 1 %%%
	\begin{squeezetable}
	\begin{longtable*}{lllddddc}
	\caption{
	\label{tab:2Ex}
	Vertical transition energies (in eV) for excited states with significant double excitation character in various molecules obtained with various methods and basis sets.
	$\%T_1$ is the percentage of single excitation calculated at the CC3 level. 
	For exFCI, an estimate of the extrapolation error is reported in parenthesis (not a statistical error bar, see text for details).
	Values from the literature are provided when available alongside their respective reference and level of theory.}
	\\
	\hline\hline
	Molecule		&	Transition					&	Method	&	\mc{4}{c}{Basis set}						&	\mcc{Lit.}	\\					
																\cline{4-7}														
					&							&			&\mcc{6-31+G(d)}&	\mcc{AVDZ}&	\mcc{AVTZ}&	\mcc{AVQZ}				\\
	\hline
	\endfirsthead
	\hline\hline
	Molecule		&	Transition					&	Method	&	\mc{4}{c}{Basis set}						&	\mcc{Lit.}	\\					
																\cline{4-7}														
					&							&			&\mcc{6-31+G(d)}&	\mcc{AVDZ}&	\mcc{AVTZ}&	\mcc{AVQZ}				\\
	\hline
	\endhead
	\hline \multicolumn{8}{r}{{Continued on next page}} \\
	\endfoot
	\hline\hline
	\multicolumn{8}{l}{$^a$Reference \onlinecite{Saha_2006}: SAC-CI results using $[4s2p1d/2s] + [2s2p2d]$ basis.}
	\\
	\multicolumn{8}{l}{$^b$Reference \onlinecite{Christiansen_1996}: CC3 results using ANO1 basis (see footnote of Table V in Ref.~\onlinecite{Christiansen_1996} for more details about the basis set).}
	\\
	\multicolumn{8}{l}{$^c$Reference \onlinecite{Barca_2018b}: Maximum overlap method (MOM) calculations at the BLYP/cc-pVTZ level.}
	\\
	\multicolumn{8}{l}{$^d$Reference \onlinecite{Galvez_2002}: Multideterminant explicitly-correlated calculations with 17 variational nonlinear parameters in the correlation factor.}
	\\
	\multicolumn{8}{l}{$^e$Reference \onlinecite{Dallos_2004}:RCA3-F/MR-CISD+Q results with aug'-cc-pVTZ.}
	\\
	\multicolumn{8}{l}{$^f$Reference \onlinecite{Watson_2012}: Incremental EOM-CC procedure (up to EOM-CCSDTQ) with CBS extrapolation.}
	\\
	\multicolumn{8}{l}{$^g$Reference \onlinecite{Chien_2018}: Heat-bath CI results using AVDZ basis.}
	\\
	\multicolumn{8}{l}{$^h$Reference \onlinecite{Boschen_2014}: CEEIS extrapolation procedure (up to sextuple excitations) with CBS extrapolation.}
	\\
	\multicolumn{8}{l}{$^i$Reference \onlinecite{Holmes_2017}: Heat-bath CI results cc-pV5Z basis.}
	\\
	\multicolumn{8}{l}{$^j$Reference \onlinecite{Barbatti_2004}: MRCISD+Q/SA3-CAS(2,2) results with AVDZ.}
	\\
	\multicolumn{8}{l}{$^k$Reference \onlinecite{Saha_2006}: SAC-CI results using $[4s2p1d/2s] + [2s2p2d]+ [2s2p]$ basis.}
	\\
	\multicolumn{8}{l}{$^l$Reference \onlinecite{Loos_2018}: exFCI/AVTZ data corrected with the difference between CC3/AVQZ and exFCI/AVTZ values.}
	\\
	\multicolumn{8}{l}{$^m$Reference \onlinecite{Angeli_2009}: State-specific PC-NEVPT2 results using ANO basis.}
	\\
	\multicolumn{8}{l}{$^n$Reference \onlinecite{Silva-Junior_2010c}: SA-CASSCF/MS-CASPT2 results using AVTZ basis.}
	\\
	\multicolumn{8}{l}{$^o$Reference \onlinecite{Schreiber_2008}: SA-CASSCF/MS-CASPT2 results using TZVP basis.}
	\endlastfoot
	Acrolein		&	1\,\ex{1}{A}{}{'} $\ra$ 3\,\ex{1}{A}{}{'}	
												&	exFCI		&	8.00(3)		&				&				&				&	8.16$^a$	\\
					&	\tr{\pi,\pi}{\pis,\pis}
												&	CC3($\%T_1$)&	8.21(73\%)	&	8.11(75\%)	&	8.08(75\%)	&				&			\\
					&							&	CASPT2		&	7.93		&	7.93		&	7.85		&	7.84		&			\\
					&							&	MS-CASPT2	&	8.36		&	8.30		&	8.28		&	8.30		&			\\
					&							&	XMS-CASPT2	&	8.18		&	8.12		&	8.07		&	8.07		&			\\
					&							&	PC-NEVPT2	&	7.91		&	7.93 		&	7.85		&	7.84		&			\\
					&							&	SC-NEVPT2	&	8.08		&	8.09 		&	8.01		&	8.00		&			\\
	\\
	Benzene			&	1\,\ex{1}{A}{1g}{} $\ra$ 1\,\ex{1}{E}{2g}{}		
												&	exFCI		&	8.40(3)		&			 	&				&				&	8.41$^b$			\\
					&	\tr{\pi,\pi}{\pis,\pis}	
												&	CCSDT		&	8.42		&	8.38	 	&				&				&		\\
					&							&	CC3($\%T_1$)&	8.50(72\%)	&	8.44(72\%) 	&	8.38(73\%)	&				&			\\
					&							&	CASPT2		&	8.43		&	8.40 		&	8.34		&	8.34		&			\\
					&							&	PC-NEVPT2	&	8.58		&	8.56 		&	8.51		&	8.52		&			\\
					&							&	SC-NEVPT2	&	8.62		&	8.61 		&	8.56		&	8.56		&			\\
	\\
					&	1\,\ex{1}{A}{1g}{} $\ra$ 2\,\ex{1}{A}{1g}{}		
												&	CASPT2		&	10.54		&	10.38 		&	10.28		&	10.27		&	10.20$^c$		\\
					&	\tr{\pi,\pi}{\pis,\pis}							
												&	MS-CASPT2	&	11.08		&	11.00 		&	10.96		&	10.97		&			\\
					&							&	XMS-CASPT2	&	10.77		&	10.64 		&	10.55		&	10.54		&			\\
					&							&	PC-NEVPT2	&	10.35		&	10.18	 	&	10.00		&				&			\\
					&							&	SC-NEVPT2	&	10.63		&	10.48 		&	10.38		&	10.36		&			\\
	\\
	Beryllium		&	1\,\ex{1}{S}{}{} $\ra$ 1\,\ex{1}{D}{}{}		
												&	exFCI		&	8.04(0)		&	7.22(0)		&	7.15(0)		&	7.11(0)		&	7.06$^d$	\\
					&	\tr{2s,2s}{2p,2p}	
												&	CCSDTQ		&	8.04		&	7.23 		&	7.15		&	7.11		&			\\
					&							&	CCSDT		&	8.04		&	7.22 		&	7.15		&	7.11		&			\\
					&							&	CC3($\%T_1$)&	8.04(2\%)	&	7.23(29\%) 	&	7.17(32\%)	&	7.12(34\%)	&			\\
					&							&	CASPT2		&	8.02	   	&	7.21		&	7.12		&	7.10		&			\\
					&							&	NEVPT2		&	8.01	   	&	7.20		&	7.11		&	7.10		&			\\
	\\
	Butadiene		&	1\,\ex{1}{A}{g}{} $\ra$ 2\,\ex{1}{A}{g}{}		
												&	exFCI		&	6.55(3)		&	6.51(12)	&				&				&	6.55$^e$, 6.39$^f$, 6.58$^g$	\\
					&	\tr{\pi,\pi}{\pi,\pi}
												&	CCSDT		&	6.63		&	6.59 		&				&				&			\\
					&							&	CC3($\%T_1$)&	6.73(74\%)	&	6.68(76\%)	&	6.67(75\%)	&	6.67(75\%)	&			\\
					&							&	CASPT2		&	6.80		&	6.78		&	6.74		&	6.75		&			\\
					&							&	PC-NEVPT2	&	6.75	   	&	6.74		&	6.70		&	6.70		&			\\
					&							&	SC-NEVPT2	&	6.83		&	6.82		&	6.78		&	6.78		&			\\
	\\
	Carbon dimer	&	1\,\ex{1}{\Sigma}{g}{+} $\ra$ 1\,\ex{1}{\Delta}{g}{}	
												&	exFCI		&	2.29(0)		&	2.21(0)		&	2.09(0)		&	2.06(0)		&	2.11$^h$	\\
					&	\tr{\pi,\pi}{\si,\si}			
												&	CCSDTQP		&	2.29		&	2.21		&				&				&			\\
					&							&	CCSDTQ		&	2.32		&	2.24 		&	2.13		&				&			\\
					&							&	CCSDT		&	2.69		&	2.63 		&	2.57		&	2.57		&			\\
					&							&	CC3($\%T_1$)&	3.10(0\%)	&	3.11(0\%) 	&	3.05(0\%)	&	3.03(0\%)	&			\\
					&							&	CASPT2		&	2.40	   	&	2.36		&	2.24		&	2.21		&			\\
					&							&	PC-NEVPT2	&	2.33	   	&	2.26		&	2.12		&	2.08		&			\\
					&							&	SC-NEVPT2	&	2.35		&	2.28 		&	2.14		&	2.11		&			\\
	\\
					&	1\,\ex{1}{\Sigma}{g}{+} $\ra$ 2\,\ex{1}{\Sigma}{g}{+}	
												&	exFCI		&	2.51(0)		&	2.50(0)		&	2.42(0)		&	2.40(0)		&	2.43$^h$, 2.46$^i$	\\
					&	\tr{\pi,\pi}{\si,\si}
												&	CCSDTQP		&	2.51		&	2.50		&				&				&			\\
					&							&	CCSDTQ		&	2.52		&	2.52 		&	2.45		&				&			\\
					&							&	CCSDT		&	2.86		&	2.87 		&	2.86		&	2.87		&			\\
					&							&	CC3($\%T_1$)&	3.23(0\%)	&	3.28(0\%) 	&	3.26(0\%)	&	3.24(0\%)	&			\\
					&							&	CASPT2		&	2.62	   	&	2.65		&	2.53		&	2.50		&			\\
					&							&	PC-NEVPT2	&	2.54	   	&	2.54		&	2.42		&	2.39		&			\\
					&							&	SC-NEVPT2	&	2.58		&	2.60 		&	2.48		&	2.44		&			\\
	\\
	Carbon trimer	&	1\,\ex{1}{\Sigma}{g}{+} $\ra$ 1\,\ex{1}{\Delta}{g}{}	
												&	exFCI		&	5.27(1)		&	5.21(0)		&	5.22(4)		&	5.23(5)		&			\\
					&	\tr{\pi,\pi}{\si,\si}
												&	CCSDTQ		&	5.35		&	5.31 		&				&				&			\\
					&							&	CCSDT		&	5.85		&	5.82 		&	5.90		&	5.92		&			\\
					&							&	CC3($\%T_1$)&	6.65(0\%)	&	6.65(0\%) 	&	6.68(1\%)	&	6.66(1\%)	&			\\
					&							&	CASPT2		&	5.13	   	&	5.06		&	5.08		&	5.08		&			\\
					&							&	PC-NEVPT2	&	5.26	   	&	5.24		&	5.25		&	5.26		&			\\
					&							&	SC-NEVPT2	&	5.21		&	5.19 		&	5.21		&	5.22		&			\\
	\\
					&	1\,\ex{1}{\Sigma}{g}{+} $\ra$ 2\,\ex{1}{\Sigma}{g}{+}	
												&	exFCI		&	5.93(1)		&	5.88(0)		&	5.91(2)		&	5.86(1)		&			\\
					&	\tr{\pi,\pi}{\si,\si}
					 							&	CCSDTQ		&	6.02		&	6.00 		&				&				&			\\
					&							&	CCSDT		&	6.52		&	6.49		&	6.57		&	6.58		&			\\
					&							&	CC3($\%T_1$)&	7.20(1\%)	&	7.20(1\%) 	&	7.24(1\%)	&	7.22(1\%)	&			\\
					&							&	CASPT2		&	5.86	   	&	5.81		&	5.82		&	5.82		&			\\
					&							&	PC-NEVPT2	&	5.97	   	&	5.97		&	5.99		&	5.99		&			\\
					&							&	SC-NEVPT2	&	5.98		&	5.97 		&	5.99		&	6.00		&			\\
	\\
	Ethylene		&	1\,\ex{1}{A}{g}{} $\ra$ 2\,\ex{1}{A}{g}{}		
												&	exFCI		&	13.38(6)	&	13.07(1)	&	12.92(6)	&				&	12.15$^j$		\\
					&	\tr{\pi,\pi}{\pis,\pis}
												&	CCSDTQP		&	13.39		&			 	&				&				&			\\
					&							&	CCSDTQ		&	13.39		&	13.07		&				&				&			\\
					&							&	CCSDT		&	13.50		&	13.20		&				&				&			\\
					&							&	CC3($\%T_1$)&	13.82(4\%)	&	13.57(15\%)	&	13.42(20\%)	&	13.06(61\%)	&			\\
					&							&	CASPT2		&	13.49		&	13.23		&	13.17		&	13.17		&			\\
					&							&	MS-CASPT2	&	13.51		&	13.26		&	13.21		&	13.21		&			\\
					&							&	XMS-CASPT2	&	13.50		&	13.25		&	13.20		&	13.20		&			\\
					&							&	PC-NEVPT2	&	14.35		&	13.42		&	13.11		&	13.04		&			\\
					&							&	SC-NEVPT2	&	13.57		&	13.33 		&	13.26		&	13.26		&			\\
	\\
	Formaldehyde	&	1\,\ex{1}{A}{1}{} $\ra$ 3\,\ex{1}{A}{1}{}		
												&	exFCI		&	10.86(1)	&	10.45(1)		&	10.35(3)		&				&	9.82$^c$	\\	
					&	\tr{n,n}{\pis,\pis}	
												&	CCSDTQP		&	10.86		&		 		&				&				&			\\
					&							&	CCSDTQ		&	10.87		&	10.44	 	&				&				&			\\
					&							&	CCSDT		&	11.10		&	10.78	 	&	10.79		&	10.80		&			\\
					&							&	CC3($\%T_1$)&	11.49(5\%)	&	11.22(4\%) 	&	11.20(5\%)	&	11.19(34\%)	&			\\
					&							&	CASPT2		&	10.80		&	10.38		&	10.27		&	10.26		&			\\
					&							&	MS-CASPT2	&	10.86		&	10.45		&	10.35		&	10.34		&			\\
					&							&	XMS-CASPT2	&	10.87		&	10.47		&	10.36		&	10.34		&			\\
					&							&	PC-NEVPT2	&	10.84		&	10.37		&	10.26		&	10.25		&			\\
					&							&	SC-NEVPT2	&	10.87		&	10.40 		&	10.30		&	10.29		&			\\
	\\
	Glyoxal			&	1\,\ex{1}{A}{g}{} $\ra$ 2\,\ex{1}{A}{g}{}		
												&	exFCI		&	5.60(1)		&	5.48(0)		&				&				&	5.66$^k$	\\
					&	\tr{n,n}{\pis,\pis}	
												&	CCSDT		&	6.24		&	6.22		&	6.35		&				&			\\
					&							&	CC3($\%T_1$)&	6.74(0\%)	&	6.70(1\%) 	&	6.76(1\%)	&	6.76(1\%)	&			\\
					&							&	CASPT2		&	5.58		&	5.47		&	5.42		&	5.43		&			\\
					&							&	PC-NEVPT2	&	5.66		&	5.56		&	5.52		&	5.52		&			\\
					&							&	SC-NEVPT2	&	5.68		&	5.58		&	5.55		&	5.55		&			\\
	\\
	Hexatriene		&	1\,\ex{1}{A}{g}{} $\ra$ 2\,\ex{1}{A}{g}{}					
%												&	exFCI		&	olympe		&			&				&				&	5.58$^g$		\\
												&	CC3($\%T_1$)&	5.78(65\%)	&	5.77(67\%)	&				&				&	5.58$^g$	\\
					&	\tr{\pi,\pi}{\pis,\pis}
												&	CCSDT		&	5.64		&	 			&				&				&			\\
					&							&	CASPT2		&	5.62		&	5.61 		&	5.58		&	5.58		&			\\
					&							&	PC-NEVPT2	&	5.67		&	5.66 		&	5.64		&	5.64		&			\\
					&							&	SC-NEVPT2	&	5.70		&	5.69		&	5.67		&	5.67		&			\\
	\\
	Nitrosomethane	&	1\,\ex{1}{A}{}{'} $\ra$ 2\,\ex{1}{A}{}{'}		
												&	exFCI		&	4.86(1)		&	4.84(2)		&	4.76(4)		&				&	4.72$^l$	\\
					&	\tr{n,n}{\pis,\pis}
												&	CCSDT		&	5.26		&	5.26 		&	5.29		&				&			\\
					&							&	CC3($\%T_1$)&	5.73(2\%)	&	5.75(4\%) 	&	5.76(3\%)	&	5.74(2\%)	&			\\
					&							&	CASPT2		&	4.93		&	4.88 		&	4.79		&	4.78		&			\\
					&							&	PC-NEVPT2	&	4.92		&	4.88 		&	4.79		&	4.78		&			\\
					&							&	SC-NEVPT2	&	4.94		&	4.90 		&	4.81		&	4.80		&			\\
	\\
	Nitroxyl		&	1\,\ex{1}{A}{}{'} $\ra$ 2\,\ex{1}{A}{}{'}		
												&	exFCI		&	4.51(0)		&	4.40(1)		&	4.33(0)		&	4.32(0)		&			\\
					&	\tr{n,n}{\pis,\pis}
												&	CCSDTQP		&	4.51		&				&				&				&			\\
					&							&	CCSDTQ		&	4.54		&	4.42		&				&				&			\\
					&							&	CCSDT		&	4.81		&	4.76		&	4.79		&	4.80		&			\\
					&							&	CC3($\%T_1$)&	5.28(0\%)	&	5.25(0\%)	&	5.26(0\%)	&	5.23(0\%)	&			\\
					&							&	CASPT2		&	4.55		&	4.46		&	4.36		&	4.34  		&			\\
					&							&	PC-NEVPT2	&	4.56		&	4.46		&	4.37		&	4.35  		&			\\
					&							&	SC-NEVPT2	&	4.58		&	4.48 		&	4.40		&	4.38		&			\\
	\\
	Pyrazine		&	1\,\ex{1}{A}{g}{} $\ra$ 2\,\ex{1}{A}{g}{}	
% 												% exFCI 		& 	17279	
												&	CC3($\%T_1$)&	9.27(7\%)	&	9.17(28\%)	&	9.17(12\%)	&				&			\\
					&	\tr{n,n}{\pis,\pis}		
					 							&	CASPT2		&	8.06		&	7.91		&	7.81		&	7.80		&			\\
					&							&	PC-NEVPT2	&	8.25		&	8.12		&	8.04		&	8.04		&			\\
					&							&	SC-NEVPT2	&	8.27		&	8.15		&	8.07		&	8.07		&			\\
	\\
					&	1\,\ex{1}{A}{g}{} $\ra$ 3\,\ex{1}{A}{g}{}		
												&	CC3($\%T_1$)&	8.88(73\%)	&	8.77(72\%)	&	8.69(71\%)	&				&			\\
					&	\tr{\pi,\pi}{\pis,\pis}	
												&	CASPT2		&	8.91		&	8.85 		&	8.77		&	8.77		&			\\
					&							&	PC-NEVPT2	&	9.12		&	9.07 		&	9.00		&	9.00		&			\\
					&							&	SC-NEVPT2	&	9.16		&	9.12 		&	9.05		&	9.05		&			\\
	\\
	Tetrazine		&	1\,\ex{1}{A}{g}{} $\ra$ 2\,\ex{1}{A}{g}{}		
%												&	exFCI		&	16904		&		 		&				&				&	4.66$^m$	\\
												&	CCSDT		&	5.86		&	5.86 		&				&				&	4.66$^m$	\\
					&	\tr{n,n}{\pis,\pis}						
												&	CC3($\%T_1$)&	6.22(1\%)	&	6.22(1\%) 	&	6.21(1\%)	&	6.19(1\%)	&			\\
					&							&	CASPT2		&	4.86		&	4.79 		&	4.69		&	4.68		&			\\
					&							&	PC-NEVPT2	&	4.75		&	4.70 		&	4.61		&	4.60		&			\\
					&							&	SC-NEVPT2	&	4.82		&	4.78 		&	4.69		&	4.68		&			\\
	\\
					&	1\,\ex{1}{A}{g}{} $\ra$ 1\,\ex{1}{B}{3g}{}		
												&	CC3($\%T_1$)&	7.64(0\%)	&	7.62(2\%)	&	7.62(1\%)	&	7.60(1\%)	&	5.76$^n$,6.01$^m$	\\
					&	\tr{n,n}{\pis_1,\pis_2}						
					 							&	CASPT2		&	6.00		&	5.95 		&	5.85		&	5.85		&			\\
					&							&	PC-NEVPT2	&	6.25		&	6.22 		&	6.15		&	6.14		&			\\
					&							&	SC-NEVPT2	&	6.30		&	6.27		&	6.20		&	6.20		&			\\
	\\
					&	1\,\ex{1}{A}{g}{} $\ra$ 1\,\ex{3}{B}{3g}{}		
												&	CC3($\%T_1$)&	7.35(5\%)	&	7.33(5\%)	&	7.35(6\%)	&	7.34(6\%)	&	5.50$^o$	\\
					&	\tr{n,n}{\pis_1,\pis_2}
					 							&	CASPT2		&	5.54		&	5.47 		&	5.39		&	5.39		&			\\
					&							&	PC-NEVPT2	&	5.63		&	5.58 		&	5.51		&	5.51		&			\\
					&							&	SC-NEVPT2	&	5.69		&	5.64 		&	5.57		&	5.57		&			\\
\end{longtable*}
\end{squeezetable}
%%% %%% %%%

%%% FIG 2 %%%
\begin{figure*}
	\begin{tabular}{ccc}
	\includegraphics[height=0.31\linewidth]{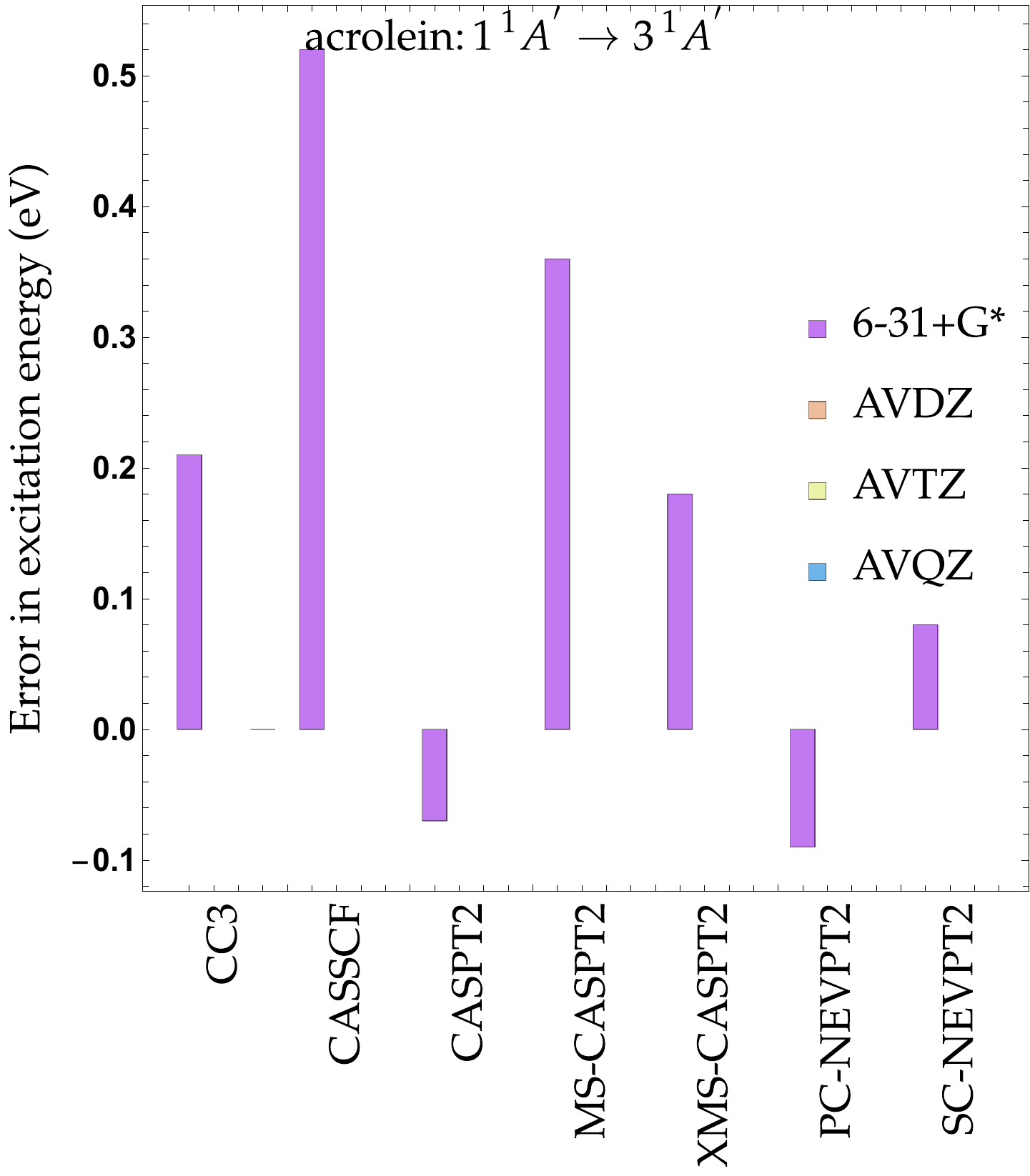}
	&
	\includegraphics[height=0.31\linewidth]{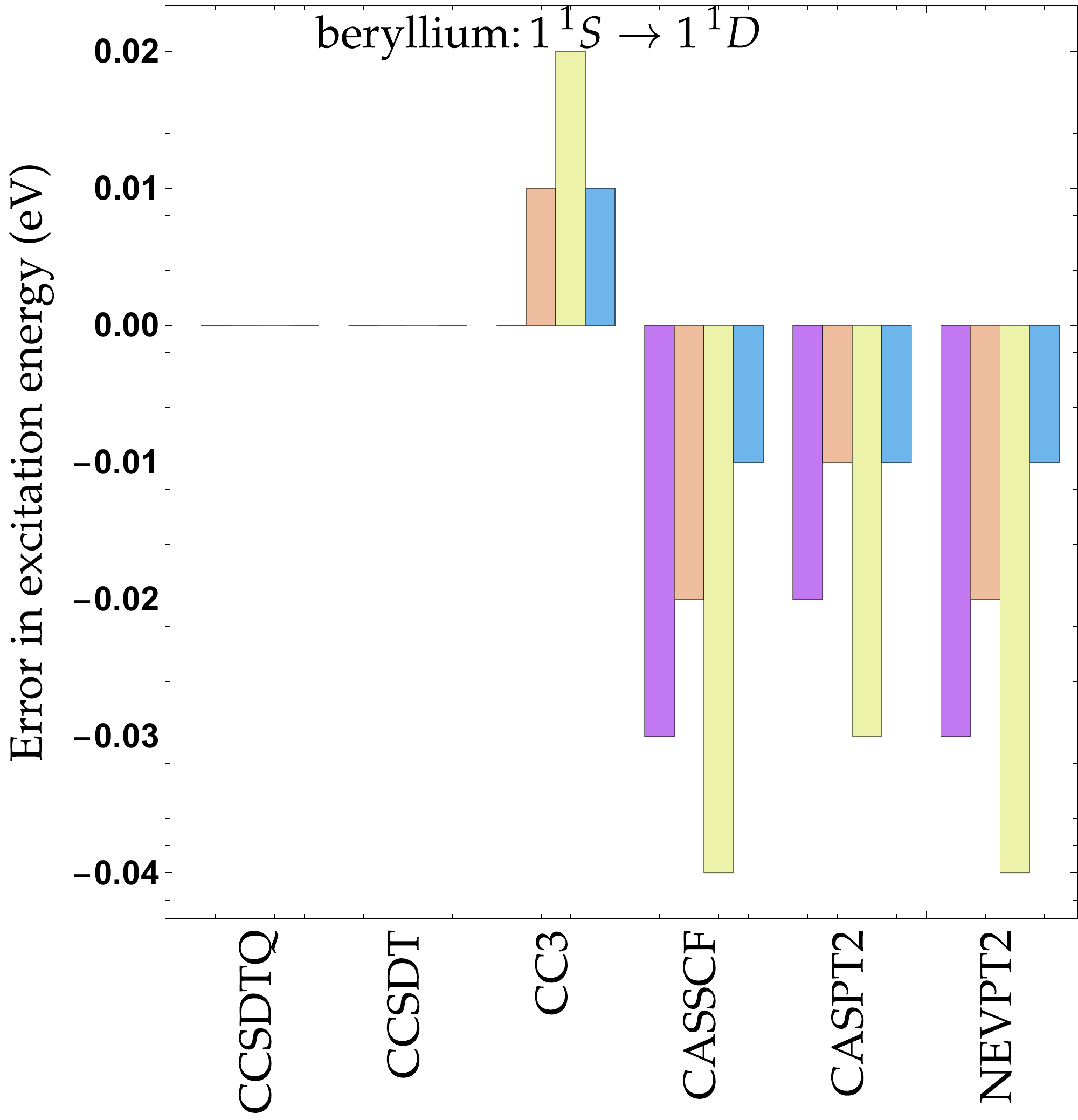}
	&
	\includegraphics[height=0.31\linewidth]{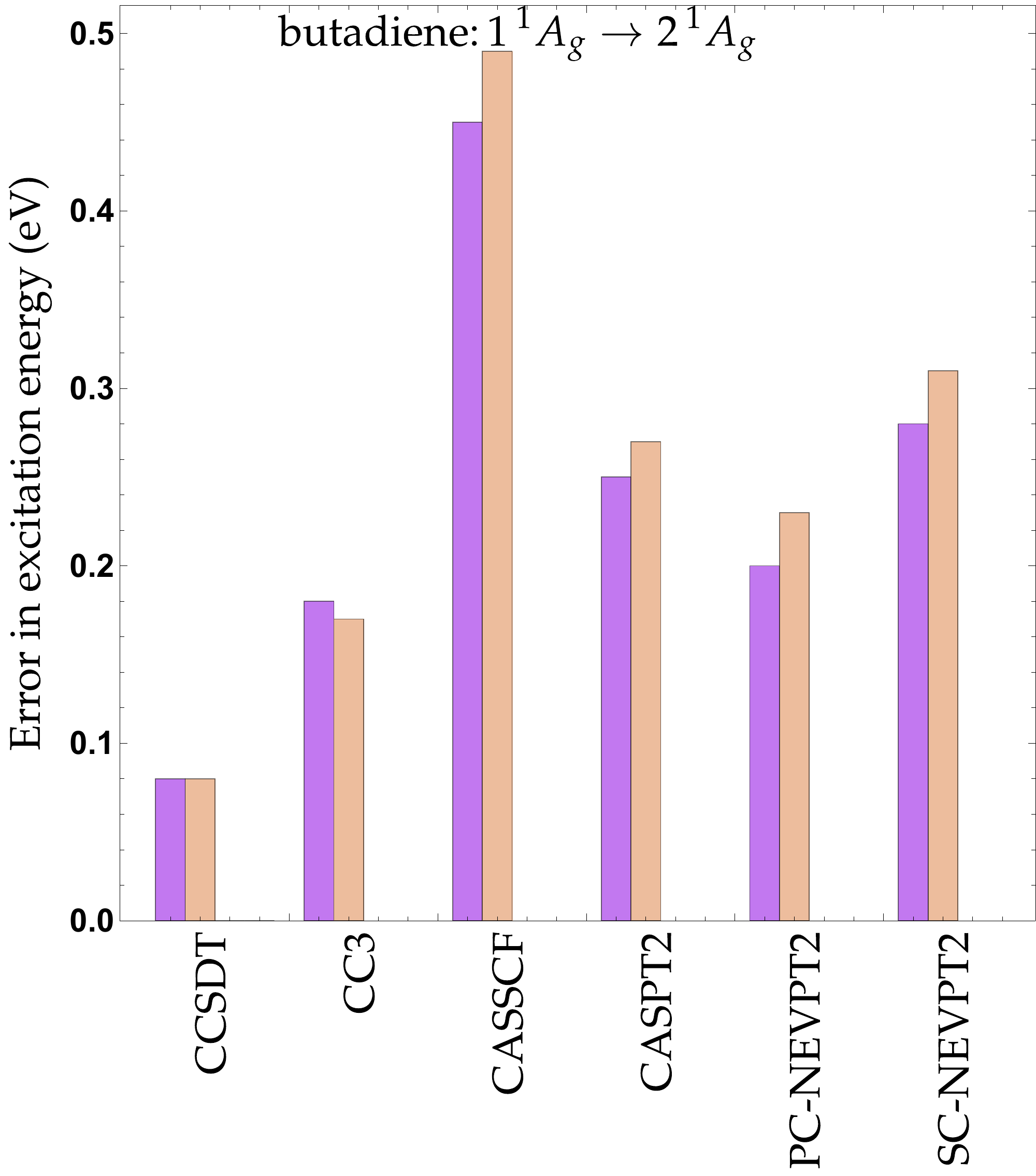}
	\\
	\includegraphics[height=0.31\linewidth]{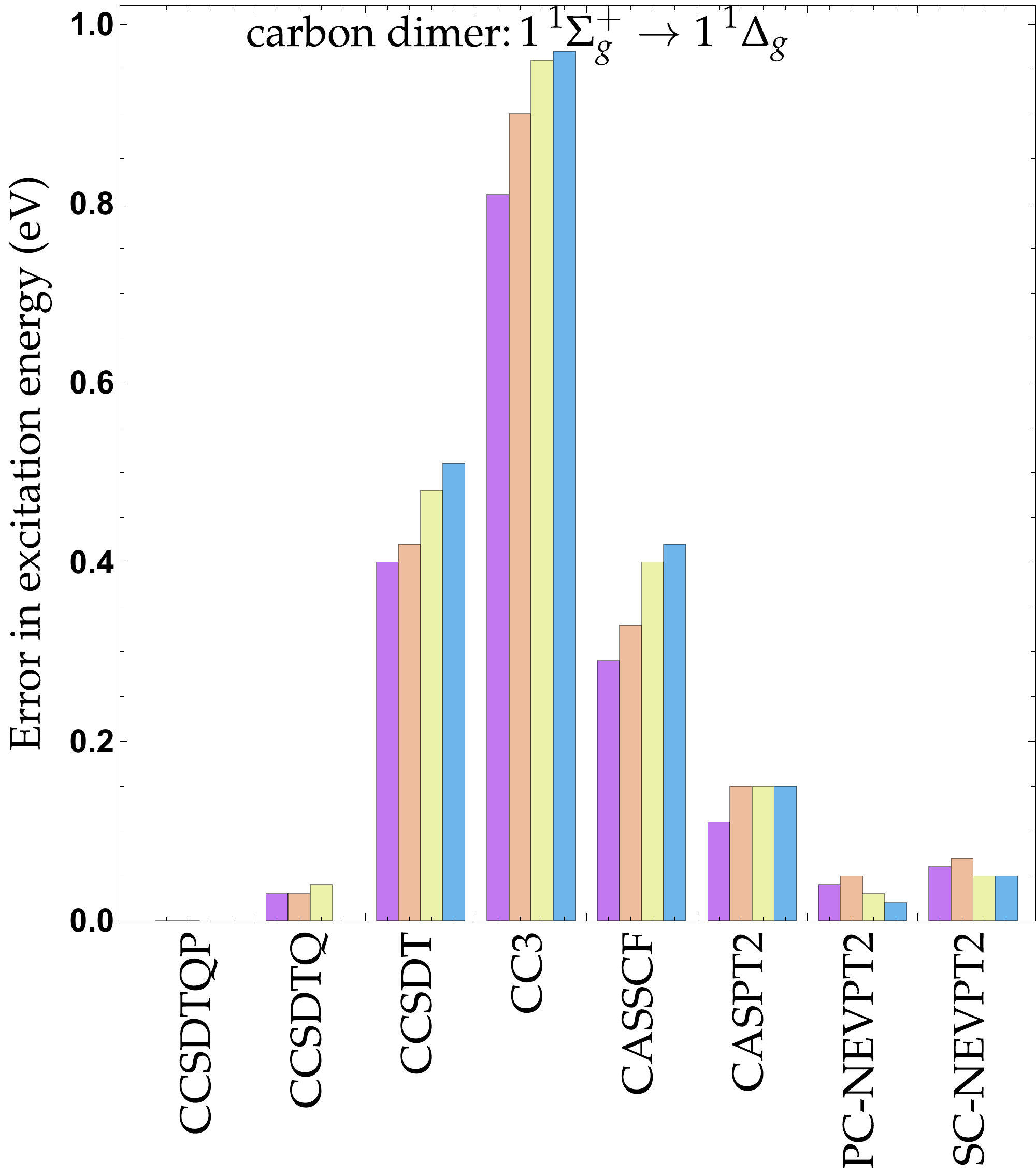}
	& 
	\includegraphics[height=0.31\linewidth]{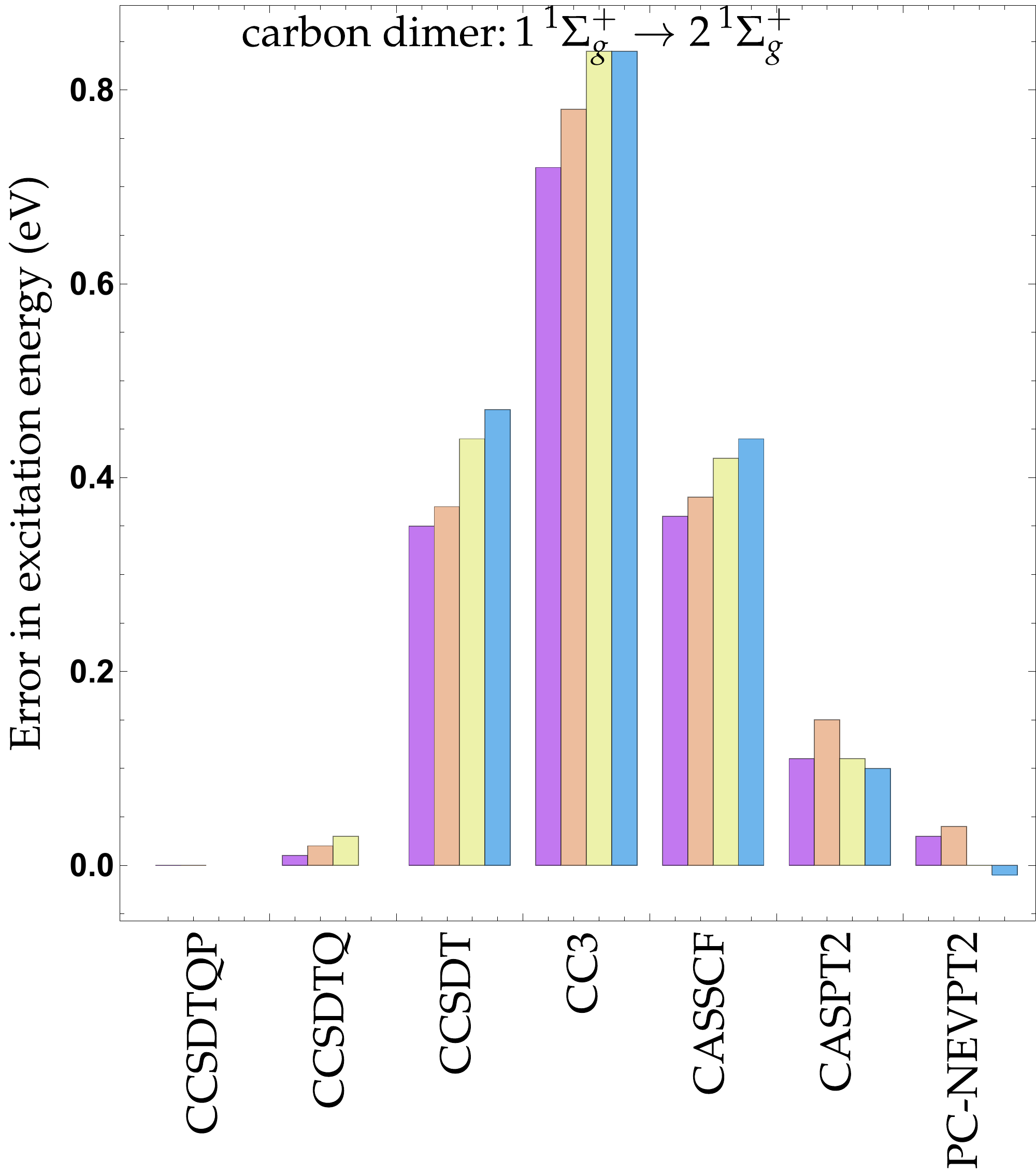}
	&
	\includegraphics[height=0.31\linewidth]{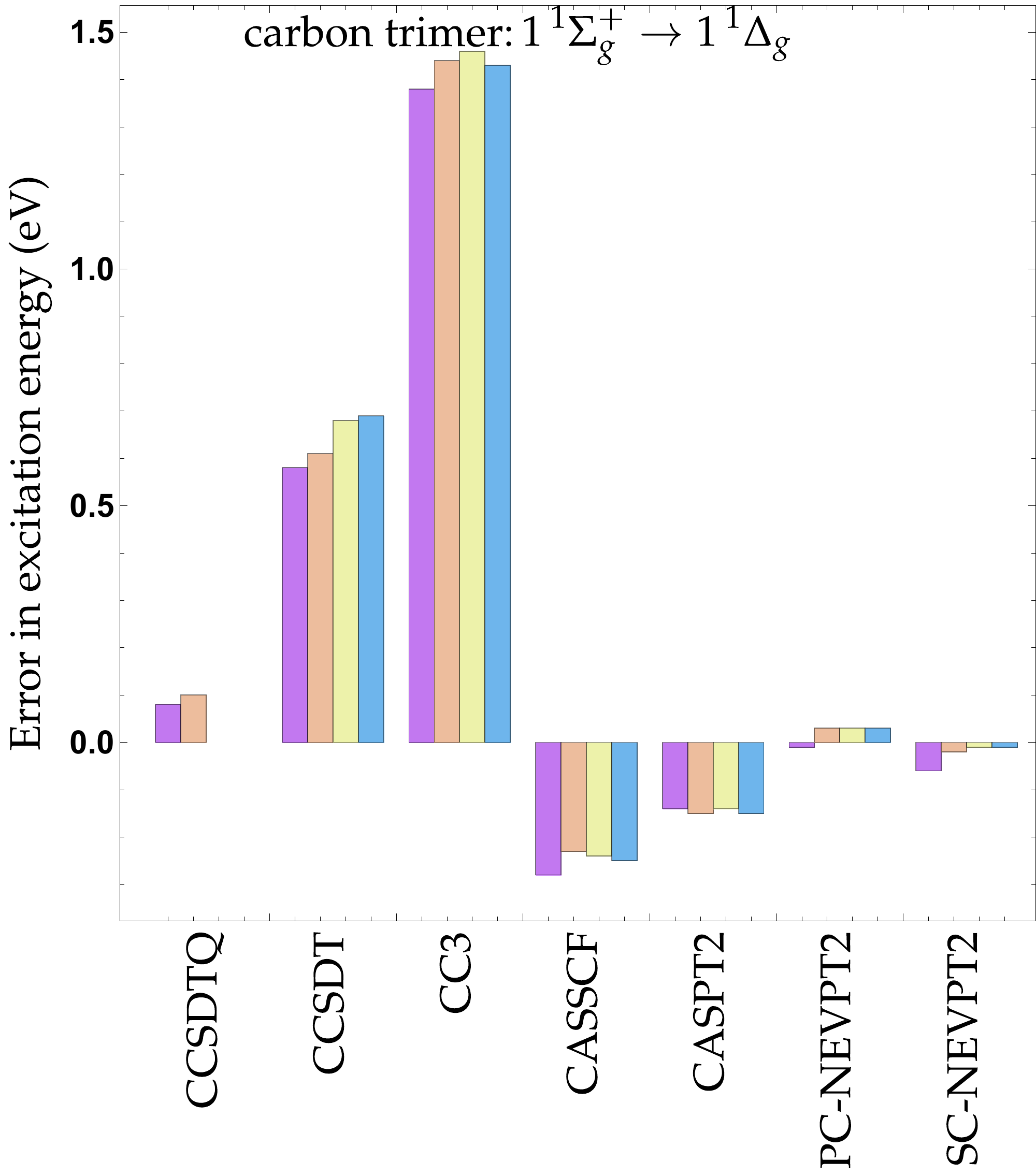}
	\\ 
	\includegraphics[height=0.31\linewidth]{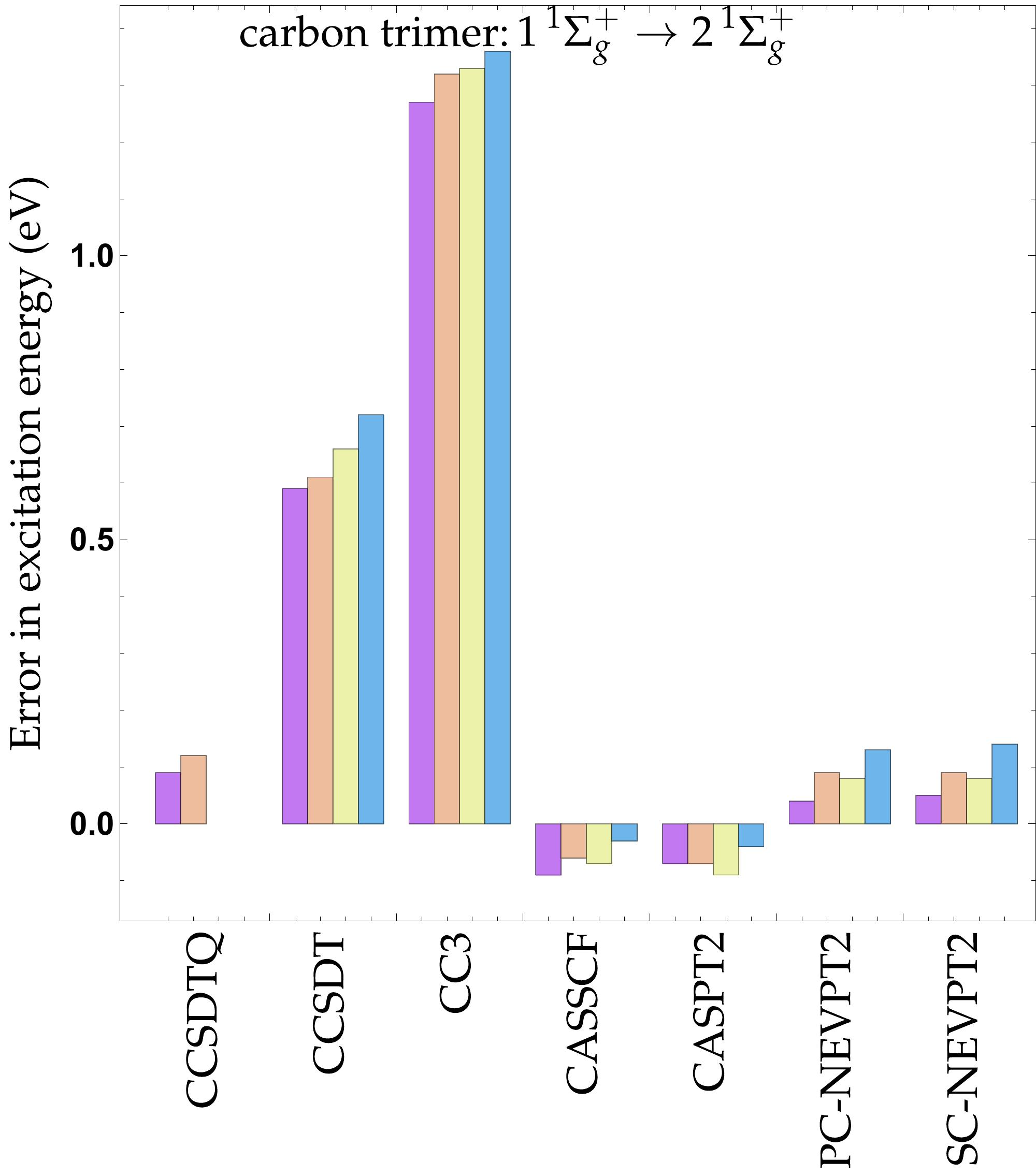}
	&
	\includegraphics[height=0.31\linewidth]{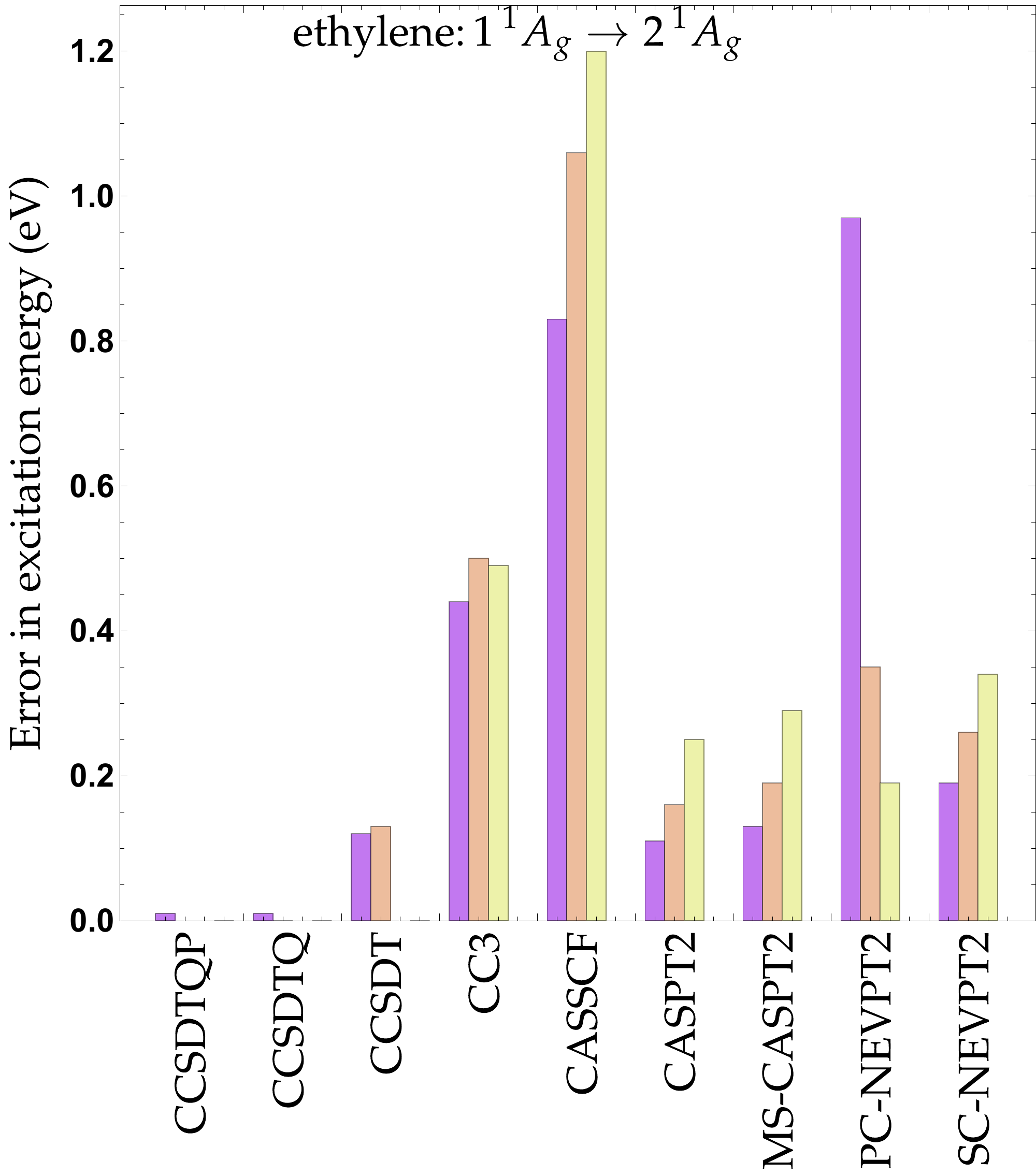}
	&
	\includegraphics[height=0.31\linewidth]{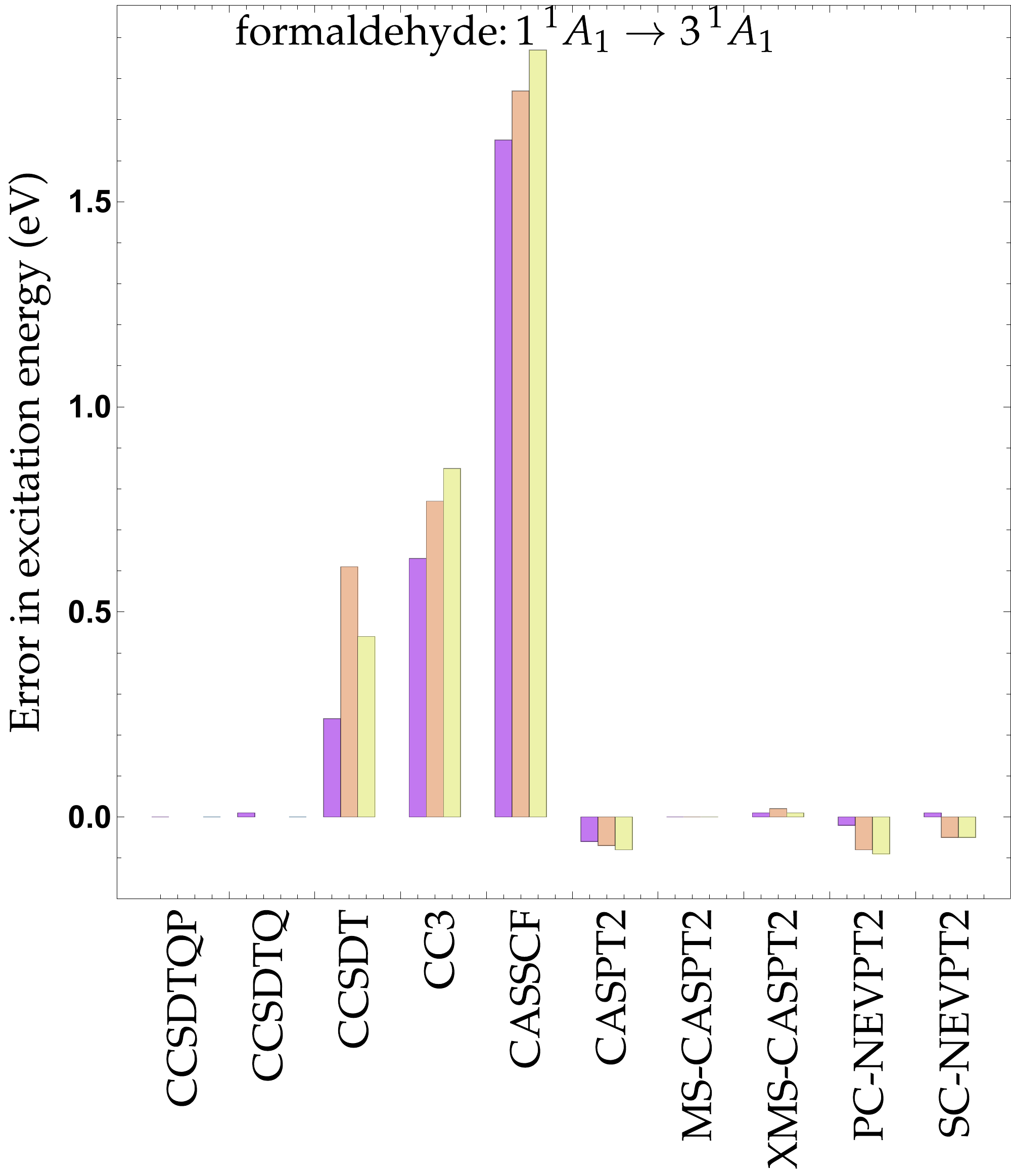}
	\\
	\includegraphics[height=0.31\linewidth]{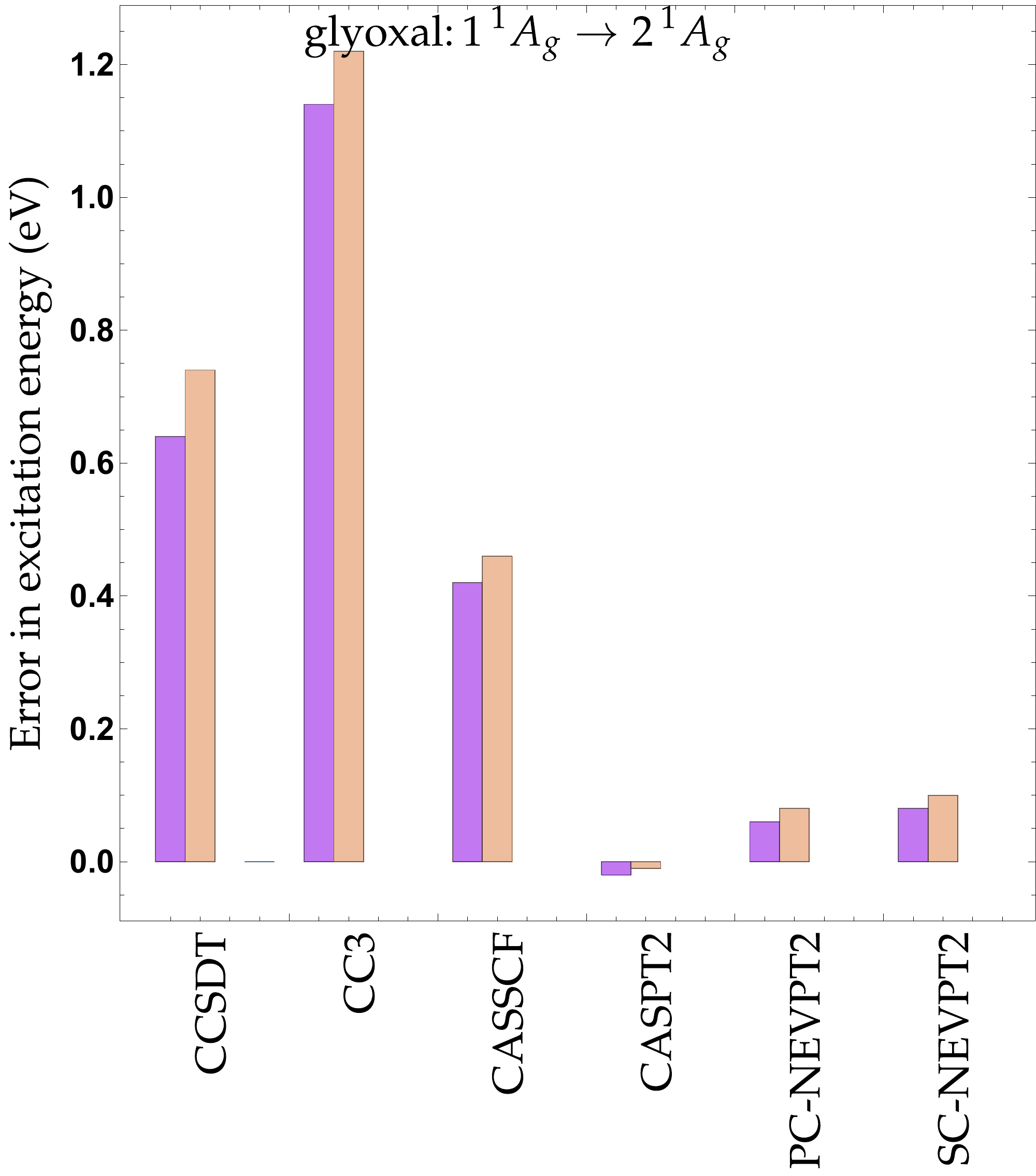}
	&
	\includegraphics[height=0.31\linewidth]{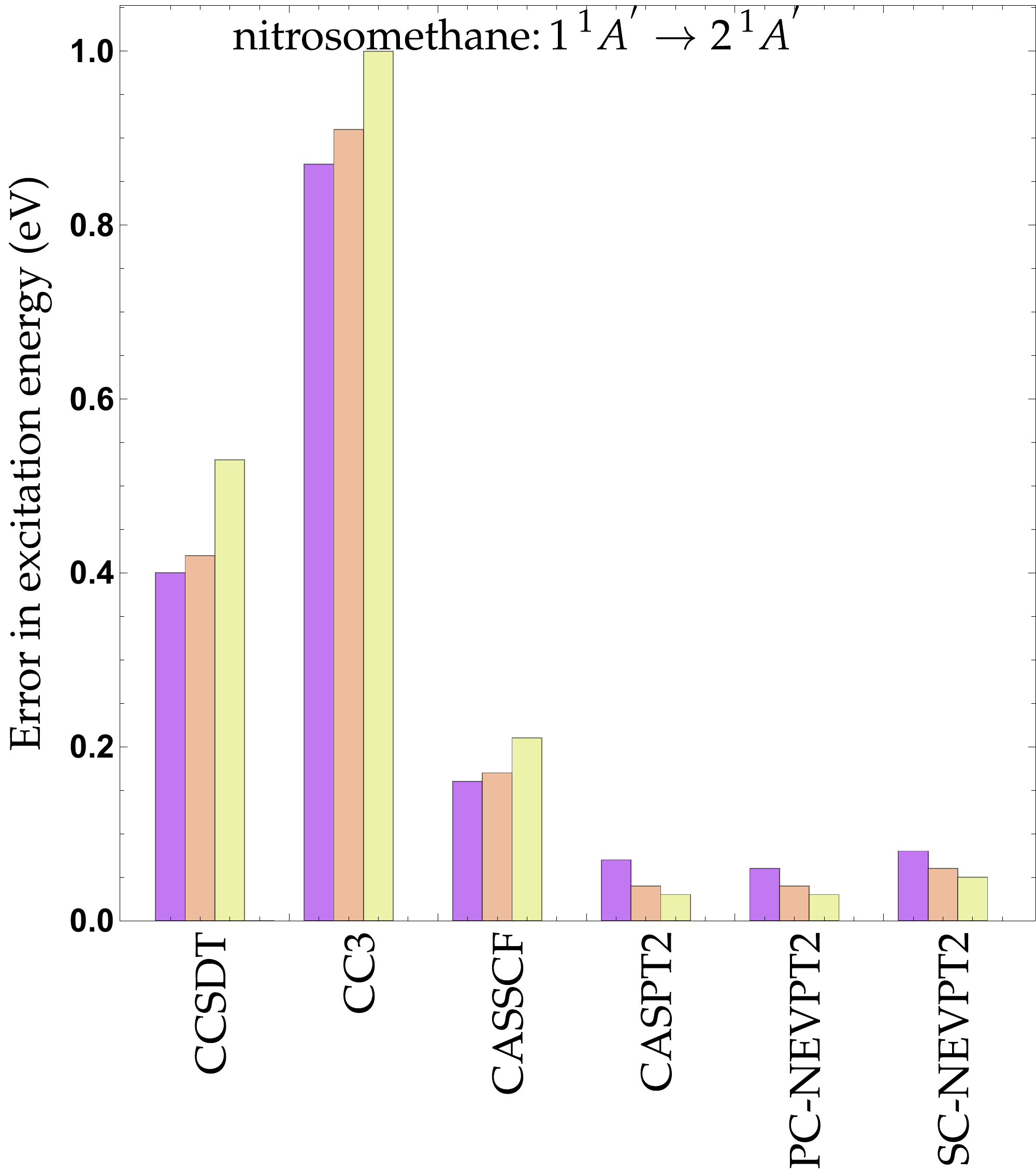}
	& 
	\includegraphics[height=0.31\linewidth]{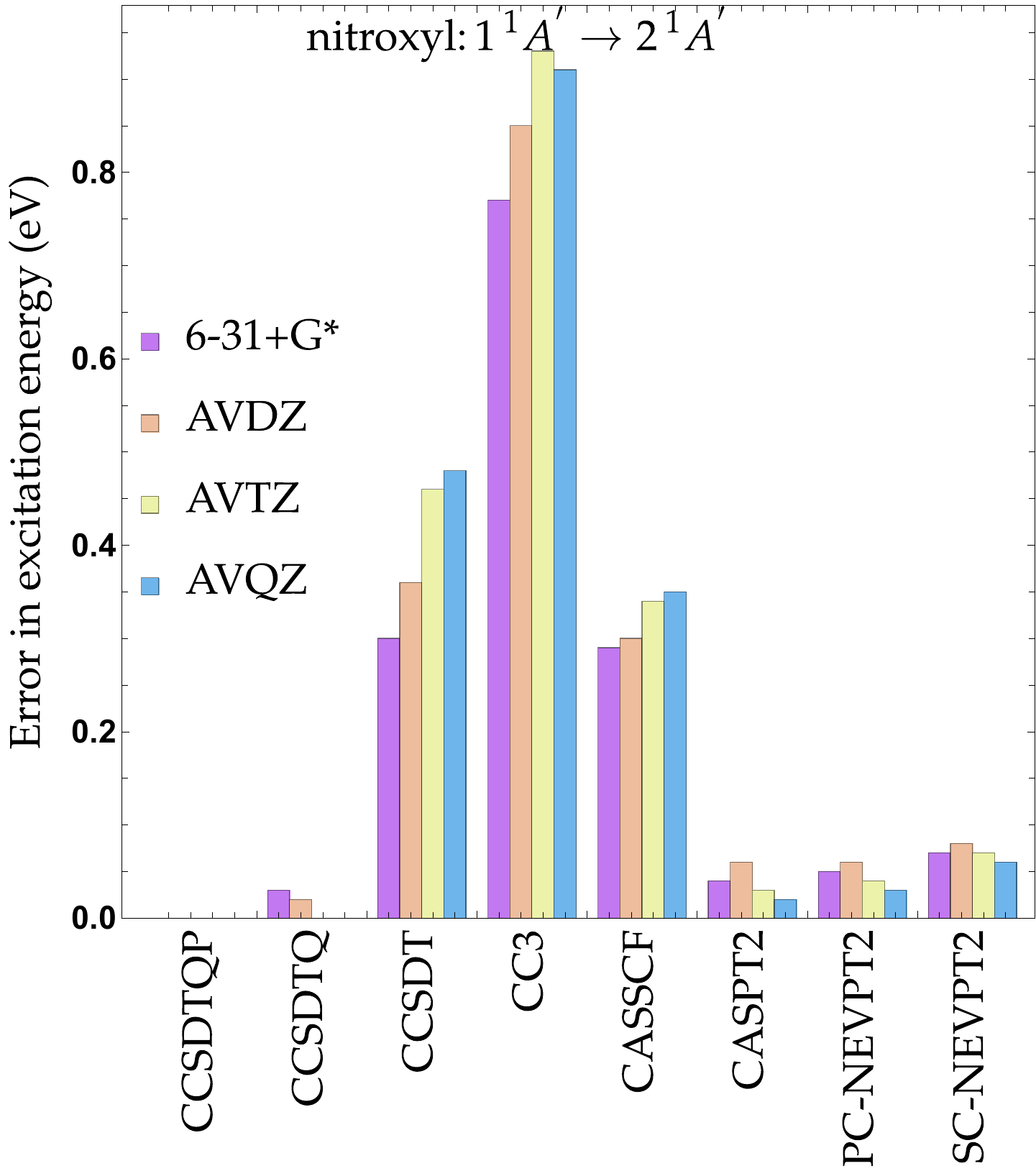}
	\\
	\end{tabular}
	\caption{
	Error in excitation energies (for a given basis and compared to exFCI) for various chemical systems, methods and basis sets.
	\label{fig:chart}
	}
\end{figure*}

%%%%%%%%%%%%%%%%%%%%%%%%
\section{
Results and discussion
\label{sec:res}
}
%%%%%%%%%%%%%%%%%%%%%%%%
The molecules considered in the present set are depicted in Fig.~\ref{fig:mol}. 
Vertical transition energies (in eV) obtained with various methods and basis sets are reported in Table \ref{tab:2Ex}, together with the nature of the transition. 
The percentage of single excitation, $\%T_1$, calculated at the CC3 level, is also reported to assess the amount of double excitation character.
Reference values taken from the literature are also reported when available.  
Total energies for each states, additional information as well as CASSCF excitation energies can be found in {\SI}. 
Finally, the error in excitation energies (for a given atomic basis set and compared to exFCI) for each system is plotted in Fig.~\ref{fig:chart}.

%---------------------------------------------
\subsection{
Beryllium
\label{sec:Be}
}
%---------------------------------------------
The beryllium atom (\ce{Be}) is the smallest system we have considered, and in this specific case, the core electrons have been correlated in all calculations.  
The lowest double excitation corresponds to the $1s^2 2s^2 (^1S) \ra 1s^2 2p^2 (^1D)$ transition.  
The $\%T_1$ values which provide an estimate of the weight of the single excitations in the CC3 calculation shows that it is mostly a double excitation with a contribution of roughly (only) $30\%$ from the singles. 

The energy of the ground and excited states of \ce{Be} have been computed by G{\`a}lves et al. \cite{Galvez_2002} using explicitly correlated wave functions, and one can extract a value of \IneV{$7.06$} for the ${}^1S \ra {}^1D$ 
transition from their study.  
This value is in good agreement with our best estimate of \IneV{$7.11$} obtained using the AVQZ basis, the difference being a consequence of the basis set incompleteness. 
Due to the small number of electrons in \ce{Be}, exFCI, CCSDT and CCSDTQ(=FCI) yield identical values for this transition for any of the basis set considered here.  
Although slightly different, the CC3 values are close to these reference values with a trifling maximum deviation of \IneV{$0.02$}.  
Irrespectively of the method, we note a significant energy difference between the results obtained with Pople's 6-31+G(d) basis and the ones obtained with Dunning's basis sets.

We have also performed multiconfigurational calculations with an active space of $2$ electrons in $12$ orbitals [CAS(2,12)] constituted by the $2s$, $2p$, $3p$ and $3d$ orbitals.
Due to the diffuse nature of the excited state, it is compulsory to take into account the $n=3$ shell to reach high accuracy.
Excitation energies computed with CASPT2 and NEVPT2 deviate by a maximum of \IneV{$0.01$} and are in excellent agreement with the exFCI numbers.

%---------------------------------------------
\subsection{
Carbon dimer and trimer
\label{sec:C2-C3}
}
%---------------------------------------------
The second system we wish to discuss is the carbon dimer (\ce{C2}) which is a prototype system for strongly correlated and multireference systems. \cite{Mulliken_1939,Clementi_1962}  Thanks to its small size, its ground and excited states 
have been previously scrutinized using highly-accurate methods. 
\cite{Abrams_2004,Sherrill_2005,Angeli_2012,Blunt_2015,Boggio-Pasqua_2000,Sharma_2015b,Booth_2011,Boschen_2014,Mahapatra_2008,Purwanto_2009a,Shi_2011,Su_2011,Toulouse_2008,Varandas_2008,Sokolov_2016a,Wouters_2014}
Here, we study two double excitations of different symmetries which are, nonetheless, close in energy: 1\,\ex{1}{\Sigma}{g}{+} $\ra$ 1\,\ex{1}{\Delta}{g}{} and  1\,\ex{1}{\Sigma}{g}{+} $\ra$ 2\,\ex{1}{\Sigma}{g}{+}. 
These two excitations --- both involving excitations from the occupied $\piCC$ orbitals to the vacant $\siCC$ orbital --- can be classified as ``pure'' double excitations, as they involve an insignificant amount of single excitations 
(see Table \ref{tab:2Ex}).  For the transition 1\,\ex{1}{\Sigma}{g}{+} $\ra$ 2\,\ex{1}{\Sigma}{g}{+}, the theoretical best estimate is most probably the \IneV{$2.46$} value reported by Holmes et al.~using the heat-bath CI method 
and the cc-pV5Z basis set at the experimental geometry. \cite{Holmes_2017}  For the 1\,\ex{1}{\Sigma}{g}{+} $\ra$ 1\,\ex{1}{\Delta}{g}{} transition, the value of \IneV{$2.11$} obtained by Boschen et al.\cite{Boschen_2014} (also at 
the experimental geometry) can be been taken as reference.  We emphasize that the value for the 1\,\ex{1}{\Sigma}{g}{+} $\ra$ 2\,\ex{1}{\Sigma}{g}{+} transition taken from this previous investigation is only \IneV{$0.03$} from the value 
reported in Ref.~\onlinecite{Holmes_2017}.

The carbon dimer constitutes a nice playground in order to illustrate the convergence of the various methods with respect to the excitation level.  For example, we have been able to perform CCSDTQP calculations for the two smallest basis sets, 
and these results perfectly agree, for each basis set, with the reference exFCI results obtained on the same (CC3) geometry.  
For all basis sets except the largest one, the CCSDTQ excitation energies are in good agreement with the exFCI results with a maximum deviation of \IneV{$0.04$}. 
With CCSDT, the error compared to exFCI ranges from $0.35$ up to half an eV, while this error keeps rising for CC3 with a deviation of the order of \IneV{$0.7$--$1.0$}.

Concerning multiconfigurational methods, we have used an active space containing 8 electrons in 8 orbitals [CAS(8,8)], which corresponds to the valence space.  
NEVPT2 is, by far, the most accurate method with errors below \IneV{$0.05$} compared to exFCI.  
As expected, the partially-contracted version of NEVPT2 yields slightly more accurate results compared to its (cheaper) strongly contracted version. CASPT2 excitation energies are consistently higher than exFCI by $0.10$--$0.15$ eV for both transitions.  
Additional calculations indicate that this bias is due to the IPEA parameter and lowering its value yields substantial improvements. 
Although CASPT2 is known to generally underestimate excitation energies for single excitations, this rule of thumb does not 
seem to apply to double excitations.

Due to its relevance in space as well as in terrestrial sooting flames and combustion processes, the carbon trimer \ce{C3} (also known as tricarbon) has motivated numerous theoretical studies. \cite{Ahmed_2004,Carter_1980,Carter_1984,Chabalowski_1986,Clementi_1962a,Hoffmann_1966,Jensen_1989,Jensen_1992,Jorgensen_1989,Kraemer_1984,Liskow_1972,Mebel_2002,Mladenovic_1994,Monninger_2002,Murrell_1990,Peric-Radic_1977,Pitzer_1959,Rocha_2015,Rocha_2016,Romelt_1978,Saha_2006b,Schroder_2016,Spirko_1997,Terentyev_2004,Varandas_2008a,Varandas_2009,Varandas_2018,Yousaf_2008}
However, its doubly-excited states have, to the best of our knowledge, never been studied.  Here, we consider the linear geometry which has been found to be the most stable isomer, although the potential energy surface 
around this minimum is known to be particularly flat. \cite{Varandas_2018}

Similarly to \ce{C2}, we have studied two transitions --- 1\,\ex{1}{\Sigma}{g}{+} $\ra$ 1\,\ex{1}{\Delta}{g}{} and  1\,\ex{1}{\Sigma}{g}{+} $\ra$ 2\,\ex{1}{\Sigma}{g}{+} --- which also both involve excitations from the occupied $\piCC$ orbitals to the vacant $\siCC$ orbitals. 
These lie higher in energy than in the dimer but remain energetically close to each other.  
Again, due to the ``pure double'' nature of the transitions, CC3 very strongly overestimates the reference values (error up to \IneV{$1.5$}).  
Interestingly, CCSDT reduces this error by roughly a factor two, bringing the deviation between CCSDT and exFCI in the  \IneV{$0.6$-$0.7$} range.  
This outcome deserves to be highlighted, as for transitions dominated by single excitations, CC3 and CCSDT have very similar accuracies as compared to exFCI. \cite{Loos_2018}  
Although very expensive, CCSDTQ brings down the error even further to a quite acceptable value of \IneV{$0.1$}.

Consistently to \ce{C2}, we have defined a $(12,12)$ active space for the trimer in order to perform multiconfigurational calculations, and we found that the CASPT2 excitation energies are consistently below exFCI by ca.~\IneV{$0.15$}. 
Again, NEVPT2 calculations are very accurate with a small preference for SC-NEVPT2, probably due to error compensation.

%---------------------------------------------
\subsection{
Nitroxyl and nitrosomethane
\label{sec:nitro}
}
%---------------------------------------------

Nitroxyl (\ce{H-N=O}) is an important molecule in biochemistry, \cite{Fukuto_2005, Miranda_2005} but only a limited number of theoretical studies of its excited states have been reported to date. \cite{Williams_1975, Luna_1995, Ehara_2011}
For this molecule, the 1\,\ex{1}{A}{}{'} $\ra$ 2\,\ex{1}{A}{}{'} transition is a genuine double excitation of $(n,n) \ra (\pis,\pis)$ nature.  
This system is small enough to perform high-order CC calculations and we have been able to push up to CCSDTQP with the 6-31+G(d) basis. 
This particular value is in perfect agreement with its exFCI analog in the same basis. For CCSDTQ, we have found that, again, the vertical excitation energies are extremely accurate, with a significant reduction of computational cost compared to CCSDTQP.  
CCSDT calculations are, as usual, significantly less accurate with an overestimation around \IneV{$0.3$}. 
CC3 adds up half an eV to this consistent overshooting of the transition energies. 

Multiconfigurational calculations have been performed with a $(12,9)$ active space corresponding to the valence space of the nitroso (\ce{-N=O}) fragment.  
In the case of nitroxyl, NEVPT2 and CASPT2 yield almost identical excitation energies, also very close to the exFCI target.

Nitrosomethane (\ce{CH3-N=O}) is an interesting test molecule \cite{Lacombe_2000, Dolgov_2004, Dolgov_2004b, Arenas_2006} and it was included in our previous study. \cite{Loos_2018} 
Similar to nitroxyl, its lowest-lying singlet $A'$ excited state corresponds to an almost pure double excitation of $(n,n) \ra (\pis,\pis)$ nature. \cite{Arenas_2006}   
Indeed, CC3/AVTZ calculations return a $3$\%\ single excitation character for this transition. Compared to nitroxyl, a clear impact of the methyl group on the double excitation energy can be noted, but overall, the same conclusions as in nitroxyl can be drawn for both CC and CAS methods. 
Therefore, we eschew discussing this case further for the sake of conciseness.
 
%---------------------------------------------
\subsection{
Ethylene and formaldehyde
\label{sec:ethyform}
}
%---------------------------------------------

Despite its small size, ethylene remains a challenging molecule that has received much attention from the theoretical chemistry community, \cite{Robin_1985, Serrano-Andres_1993, Watts_1996, Schreiber_2008, Angeli_2008, Feller_2014, Chien_2018} 
and is included in many benchmark sets. \cite{Head-Gordon_1994, Schreiber_2008, Shen_2009b, Caricato_2010, Leang_2012, Hoyer_2016, Loos_2018}   In particular, we refer the interested readers to the work of Davidson and coworkers\cite{Feller_2014} 
for, what we believe, is the most complete and accurate investigation dedicated to the excited states of ethylene. 

In ethylene, the double excitation 1\,\ex{1}{A}{g}{} $\ra$ 2\,\ex{1}{A}{g}{} is of \tr{\pi,\pi}{\pis,\pis} nature.  Unsurprisingly, it has been much less studied than the single excited states due to its fairly high energy and the absence of experimental 
value.  Nevertheless, in 2004, Barbatti et al.~have reported a value of \IneV{$12.15$} at the MRCISD+Q/AVDZ level of theory. \cite{Barbatti_2004} We have found that this state has a fairly high degree of double excitation which, at the CC3 level, 
decreases with the size of the basis set, with $\%T_1$ going from $4\%$ with 6-31+G(d)  to $61\%$ with AVQZ. Due to its Rydberg character, there is obviously a large basis set effect for this transition, with a magnitude that is additionally strongly 
method dependent.

Here again, thanks to the small size of this molecule, we have been able to perform high-order CC calculations, and, once more, we have found that CCSDTQP and CCSDTQ yield very accurate excitation energies. 
Removing the quadruples has the effect of blue-shifting the transition by at least \IneV{0.1}, while CC3 is off by half an eV independently of the basis set.

In the case of ethylene, we have studied two types of active spaces: a $(2,2)$ active space which includes the $\piCC$ and $\pisCC$ orbitals and a $(4,4)$ active space obtained by adding the $\siCC$ and $\sisCC$ orbitals.
Table \ref{tab:2Ex} only reports the results for the largest active space; the values determined with the smallest active space can be found in {\SI}.  In accordance with previous studies, \cite{Angeli_2008, Feller_2014, Giner_2015b} we have 
found that it is essential to take into account the bonding and anti-bonding $\si$ orbitals in the active space due to the strong coupling between the $\si$ and $\pi$ spaces.  CASPT2 and NEVPT2 are overestimating the transition energy 
by at least \IneV{$0.2$} with Dunning's bases, while CASPT2 and MS-CASPT2 yield similar excitation energies.  We note that the PC-NEVPT2 energies seem to become more accurate when the quality of the atomic basis 
set improves, whereas the opposite trend is observed for SC-NEVPT2.

From a computational point of view, formaldehyde is similar to ethylene and it has also been extensively studied at various levels of theory. 
\cite{Merchan_1995, Paterson_2006, Muller_2001, Foresman_1992b, Hadad_1993, Head-Gordon_1994, Head-Gordon_1995, Gwaltney_1995, Wiberg_1998, Wiberg_2002, Peach_2008, Schreiber_2008, Shen_2009b, Caricato_2010, Li_2011, Leang_2012, Hoyer_2016,Loos_2018} 
However, the 1\,\ex{1}{A}{1}{} $\ra$ 3\,\ex{1}{A}{1}{} transition in \ce{CH2=O} is rather chemically different from its \ce{H2C=CH2} counterpart, as it is a transition from the ground state to the second excited state of \ex{1}{A}{1}{} symmetry with 
a \tr{n,n}{\pis,\pis} character.  For this transition, Barca et al.\cite{Barca_2018a} have reported a value of \IneV{$9.82$} at the BLYP/cc-pVTZ level [using the maximum overlap method (MOM) to locate the excited state] in qualitative agreement with our reference 
energies. The lack of diffuse functions may have, however, a substantial effect on this value. 

In terms of the performance of the CC-based methods, the conclusion that we have drawn in ethylene can be almost perfectly transposed to formaldehyde. For the CAS-type calculations, two active spaces were tested: a $(4,3)$ active space 
that includes the  $\piCO$ and $\pisCO$ orbitals as well as the lone pair $\nO$ on the oxygen atom, and the $(6,5)$ active space that adds the $\siCO$ and $\sisCO$ orbitals.  
Again, Table \ref{tab:2Ex} only reports the results obtained with the largest  active space whereas the values for the smallest active space can be found in {\SI}.  
The performance of multiconfigurational calculations are fairly consistent and there are no significant differences between the various methods, although, due to the strong mixing between the first three \ex{1}{A}{1}{} states, the results obtained with CASPT2, MS-CASPT2, and XMS-CASPT2 differ slightly. 
The excitation energies obtained with the multi-state variants (extended or not) almost perfectly match the exFCI values, thanks to a small blueshift of the energies compared to the CASPT2 results.  
Note that the same methods would return excitation energies with errors consistently red-shifted by \IneV{$0.15$} with the small active space, highlighting once more that $\sigma$ orbitals should be included if high accuracy is desired.

%---------------------------------------------
\subsection{
Butadiene, glyoxal, and acrolein
\label{sec:butadiene}
}
%---------------------------------------------

The excited states of (\emph{trans}-)butadiene have been thoroughly studied during the past thirty years.
\cite{Watts_1996, Boggio-Pasqua_2004, Cave_2004, Chien_2018, Daday_2012, Dallos_2004, Hsu_2001, Kitao_1988, McDiarmid_1988, Mosher_1973, Ostojic_2001, Saha_2006, Serrano-Andres_1993, Strodel_2002, Watson_2012, Sundstrom_2014, Shu_2017}
In 2012, Watson and Chan \cite{Watson_2012} have studied the hallmark singlet bright (1\,\ex{1}{B}{u}{}) and dark (2\,\ex{1}{A}{g}{}) states. They reported best estimates of $6.21 \pm 0.02$ eV and $6.39 \pm 0.07$ eV, respectively, settling 
down the controversy about the ordering of these two states. \cite{Strodel_2002} While the bright 1\,\ex{1}{B}{u}{} state has a clear (HOMO $\ra$ LUMO) single excitation character, the dark 2\,\ex{1}{A}{g}{} state includes a substantial fraction 
of doubly-excited character from the HOMO $\ra$ LUMO double excitation (roughly $30\%$), yet dominant contributions from the HOMO-1$\ra$LUMO and HOMO$\ra$LUMO+1 single excitations.  Butadiene (as well as hexatriene, see below) has 
been also studied at the dressed TD-DFT level. \cite{Maitra_2004, Cave_2004, Huix-Rotllant_2011}

For butadiene (and the two other molecules considered in this Section), exFCI results are only reported for the two double-$\zeta$ basis sets, as it was not possible to converge the excitation energies with larger basis sets. Our exFCI estimates agree nicely with the 
reference values obtained by Dallos and Lischka, \cite{Dallos_2004} Watson and Chan, \cite{Watson_2012} and Chien et al.\cite{Chien_2018} at the MR-CI, incremental CC and heat-bath CI levels, respectively (see Table \ref{tab:2Ex}).

Concerning the multiconfigurational calculations, the $(4,4)$ active space includes the $\piCC$ and $\pisCC$ orbitals while the $(10,10)$ active space adds the $\siCC$ and $\sisCC$ orbitals.  Expanding the active space has a non-negligible
 impact on the NEVPT2 excitation energies with a neat improvement by ca.~$0.1$ eV, whereas CASPT2 results are less sensitive to this  active space expansion  (see {\SI}).  As previously mentioned, this effect is reminiscent of the strong 
 coupling between the $\si$ and $\pi$ spaces in compounds like butadiene, \cite{Watson_2012, Dash_2018} ethylene, \cite{Angeli_2008, Giner_2015b} and cyanines. \cite{Send_2011, Boulanger_2014, LeGuennic_2015, Garniron_2018}
Here, it is important to note that both CC3 and CCSDT provide more accurate excitation energies than any multiconfigurational method.  This clearly illustrates the strength of CC approaches when there is a dominant ``single'' nature in the 
considered transition as discussed in previous works. \cite{Shu_2017, Barca_2018a, Barca_2018b}

The genuine double excitation 1\,\ex{1}{A}{g}{} $\ra$ 2\,\ex{1}{A}{g}{} in glyoxal, \cite{Dykstra_1977, Ha_1972, Hirao_1983, Hollauer_1991, Pincellt_1971, Saha_2006} which corresponds to a \tr{n,n}{\pis,\pis} transition, has been studied by Saha et al.~at the SAC-CI level \cite{Saha_2006} (see Table \ref{tab:2Ex} for additional information). They reported a value of \IneV{$5.66$} in very good agreement with our exFCI reference. 
As expected now, given the ``pure'' double excitation character, CC3 and CCSDT are off by the usual margin (more than \IneV{$1$} for CC3). Due to the nature of the considered transition, the lone pairs of the two oxygen atoms are included in both the small $(8,6)$ and large $(14,12)$ active spaces. 
In glyoxal, we have logically found that the lone pairs of both oxygen atoms  equally contribute to the double excitation. 
The $(8,6)$ active space also contains the $\piCC$, $\piCO$, $\pisCC$ and $\pisCO$ orbitals, while the $(14,12)$ active space adds up the $\siCC$, $\siCO$, $\sisCC$ and $\sisCO$ orbitals.
CASPT2 excitation energies are particularly close to our exFCI energies while PC- and SC-NEVPT2 energies are slightly blue-shifted but remain in very good agreement with the exFCI benchmark.

The 1\,\ex{1}{A}{}{'} $\ra$ 3\,\ex{1}{A}{}{'} excitation in acrolein \cite{Aquilante_2003, Bouabca_2009, Guareschi_2013, Saha_2006} has the same nature as the one in butadiene.   
However, there is a 1\,\ex{1}{A}{}{'} $\ra$ 2\,\ex{1}{A}{}{'} transition of $\pi \ra \pis$ nature slightly below in energy and these two transitions are strongly coupled.  From a computational point of view, it means that the 1\,\ex{1}{A}{}{'} $\ra$ 3\,\ex{1}{A}{}{'} transition is, from a technical point of view, tricky to get, and this explains why we have not been able to obtain reliable exFCI estimates except for the smallest 6-31+G(d) basis.

The (small) $(4,4)$ active space contains the $\piCC$, $\piCO$, $\pisCC$ and $\pisCO$ orbitals, while the (larger) $(10,10)$ active space adds up the $\siCC$, $\siCO$, $\sisCC$ and $\sisCO$ orbitals. 
Due to the nature of the transitions involved, it was not necessary to include the lone pair of the oxygen atom in the active space, and this has been confirmed by preliminary calculations. 
Moreover, CASSCF predicts the $\pi \ra \pis$ transition higher in energy than the \tr{\pi,\pi}{\pis,\pis} transition, and CASPT2 and NEVPT2 correct this erroneous ordering via the introduction of dynamic correlation.
The CAS(4,4) calculations clearly show that the multi-state treatment of CASPT2 strongly mix these two transitions, while its extended variant mitigates this trend.
Consequently, because of the strong mixing of the three \ex{1}{A}{}{'} states in acrolein, CASPT2, MS-CASPT2 and XMS-CASPT2 deviate by several tenths of eV. 

For the 1\,\ex{1}{A}{}{'} $\ra$ 3\,\ex{1}{A}{}{'} excitation of acrolein, Saha et al.\cite{Saha_2006} provided an estimate of \IneV{$8.16$} at the SAC-CI level as compared to our exFCI/6-31+G(d) value of \IneV{$8.00$}, which nestles between the PC- and SC-NEVPT2 values.
The CC3 excitation energy in the same basis is off by ca.~\IneV{$0.2$}, so is the XMS-CASPT2 energy.

%---------------------------------------------
\subsection{
Benzene, pyrazine, tetrazine, and hexatriene
\label{sec:BigMol}
}
%---------------------------------------------
In this last section, we report excitation energies for four larger molecules containing 6 heavy atoms (see Fig.~\ref{fig:mol}). 
Due to their size, we have not been able to provide reliable exFCI results (except for benzene, see below). 
Therefore, we mainly restrict ourselves to multiconfigurational calculations with valence $\pi$ active space as well as with CC3 and CCSDT (when technically possible). 
For the nitrogen-containing molecules, the lone pairs have been included in the active space as we have found that they are always involved in double excitations. 
We refer the reader to the {\SI} for details about the active spaces.

Thanks to the high degree of symmetry of benzene, we have been able to obtain a reliable estimate of the excitation energy at the exFCI/6-31+G(d) for the lowest double excitation  of 1\,\ex{1}{A}{1g}{} $\ra$ 1\,\ex{1}{E}{2g}{}
character. \cite{Buenker_1968, Peyerimhoff_1970, Hay_1974, Palmer_1989, Kitao_1987, Lorentzon_1995, Hashimoto_1996, Christiansen_1996, Christiansen_1998, Handy_1999, Hald_2002, Barca_2018a, Barca_2018b} 
Our value of \IneV{$8.40$} is in almost perfect agreement with the one reported by Christiansen et al.\cite{Christiansen_1996} at the CC3 level (\IneV{$8.41$}). 
Indeed, as this particular transition has a rather small double excitation character, CC3 and CCSDT provide high-quality results. 
This contrasts with the 1\,\ex{1}{A}{1g}{} $\ra$ 2\,\ex{1}{A}{1g}{} transition which has almost a pure double excitation nature.  
This genuine double excitation has received less attention but Gill and coworkers reported a value of \IneV{$10.20$} at the BLYP(MOM)/cc-pVTZ level in nice agreement with our CASPT2 results.
However, we observe that depending on the flavor of post-CASSCF treatment, we have an important variation (by ca.~\IneV{$0.6$--$0.9$}) of the excitation energies, the lower and upper bounds being respectively provided by PC-NEVPT2 and MS-CASPT2.

For pyrazine, \cite{Palmer_1991, Fulscher_1992, Durig_1984, Flscher_1994, Weber_1999, Nooijen_1999} we have studied the three lowest states of \ex{1}{A}{g}{} symmetry and their corresponding excitation energies.
The 1\,\ex{1}{A}{g}{} $\ra$ 2\,\ex{1}{A}{g}{} transition of \tr{n,n}{\pis,\pis} nature has a large fraction of double excitation, while the 1\,\ex{1}{A}{g}{} $\ra$ 3\,\ex{1}{A}{g}{} transition has a \tr{\pi,\pi}{\pis,\pis} nature, and is dominated by single excitations, similar to the one studied in butadiene and acrolein. 
In pyrazine, both lone pairs contribute to the second excitation.
One can note an interesting methodological inversion between these two transitions. Indeed, due to the contrasted quality of CC3 excitation energies for the \tr{n,n}{\pis,\pis} and \tr{\pi,\pi}{\pis,\pis} transitions, the latter 
is (incorrectly) found below the former at the CC3 level while the opposite is observed with CASPT2 or NEVPT2.

Tetrazine (or s-tetrazine) \cite{Mason_1959, Innes_1988, Fridh_1972, Palmer_1997, Livak_1971, Rubio_1999, Nooijen_2000, Schreiber_2008, Harbach_2014} is a particularly ``rich'' molecule in terms of double excitations thanks to the 
presence of four lone pairs. Here, we have studied three transitions: two singlet-singlet and one singlet-triplet excitations. In these three transitions, electrons from the nitrogen lone pairs $\nN$ are excited to $\pis$ orbitals.
As expected, they can be labeled as genuine double excitations as they have very small $\%T_1$ values. 
For the 1\,\ex{1}{A}{g}{} $\ra$ 1\,\ex{1}{B}{3g}{} and 1\,\ex{1}{A}{g}{} $\ra$ 1\,\ex{3}{B}{3g}{} transitions, we note that the two 
excited electrons end up in different $\pis$ orbitals, contrary to most cases encountered in the present study. The basis set effect is pretty much inexistent for these three excitations with a maximum difference of \IneV{$0.04$} between 
the smallest and the largest basis sets. For tetrazine, previous high-accuracy reference values are: i) \IneV{$4.66$} for the 1\,\ex{1}{A}{g}{} $\ra$ 2\,\ex{1}{A}{g}{} transition reported by Angeli et al.\cite{Angeli_2009} with 
NEVPT2, ii) \IneV{$5.76$} for the 1\,\ex{1}{A}{g}{} $\ra$ 1\,\ex{1}{B}{3g}{} transition reported by Silva-Junior et al. \cite{Silva-Junior_2010c} at the MS-CASPT2/AVTZ level and, iii) \IneV{$5.50$} for the 1\,\ex{1}{A}{g}{} $\ra$ 1\,\ex{3}{B}{3g}{} 
transition reported by Schreiber et al. \cite{Schreiber_2008} at the MS-CASPT2/TZVP level. In comparison, for the second transition, Angeli et al.\cite{Angeli_2009} have obtained a value of \IneV{$6.01$} at the NEVPT2 level.
For the first transition, the CCSDT results indicate that the CC3 excitation energies are, again, fairly inaccurate and pushing up to CCSDT does not seem to  significantly improve the results as the deviations between CCSDT and CASPT2/NEVPT2 results are still substantial.
However, it is hard to determine which method is the most reliable in this case.
Finally, we note that, for the second and third transitions, there is an important gap between CASPT2 and NEVPT2 energies.

For hexatriene, \cite{Serrano-Andres_1993, Flicker_1977, Nakayama_1998, Maitra_2004, Cave_2004} the accurate energy of the 2\,\ex{1}{A}{g}{} state is not known experimentally, illustrating the difficulty to observe these states via conventional spectroscopy techniques. 
For this molecule, we have unfortunately not been able to provide reliable exFCI results, even for the smallest basis sets. 
However, Chien et al.~have recently reported a value of \IneV{5.58} at the heat-bath CI/AVDZ level with a MP2/cc-pVQZ geometry. \cite{Chien_2018}  
This reference value indicates that our CASPT2 and NEVPT2 calculations are particularly accurate even with a minimal valence $\pi$ active space, the coupling between $\si$ and $\pi$ spaces becoming weaker for larger polyenes. \cite{Garniron_2018}
Because the 1\,\ex{1}{A}{g}{} $\ra$ 2\,\ex{1}{A}{g}{} transition is of \tr{\pi,\pi}{\pis,\pis} nature (and very similar to its butadiene analog), the CC3 transition energies are not far off the reference values.

%%% TABLE 2 %%%
\begin{squeezetable}
\begin{table*}
	\caption{
	\label{tab:TBE}
	Theoretical best estimates (TBEs) of vertical transition energies (in eV) for excited states with significant double excitation character in various molecules (see Table \ref{tab:2Ex} for details).
	TBEs are computed as $\Delta E_\text{R/SB} + \Delta E_\text{C/LB} - \Delta E_\text{C/SB}$, where $\Delta E_\text{R/SB}$ is the excitation energy computed with a reference (R) method in a small basis (SB), and $\Delta E_\text{C/SB}$ and $\Delta E_\text{C/LB}$ are excitation energies computed with a correction (C) method in the small and large basis (LB), respectively. 
	}
	\begin{ruledtabular}
	\begin{tabular}{llldldd}
	Molecule		&	Transition												&	\mc{2}{c}{Reference}			&	\mc{2}{c}{Correction}				&	\mcc{TBE}	\\
																				\cline{3-4}							\cline{5-6}
					&															&	Level R/SB						&	\mcc{$\Delta E_\text{R/SB}$}			
																				&	Level C/LB						&	\mcc{$\Delta E_\text{C/LB} - \Delta E_\text{C/SB}$}				\\
	\hline
	Acrolein		&	1\,\ex{1}{A}{}{'} $\ra$ 3\,\ex{1}{A}{}{'}				&	exFCI/6-31+G(d)					&	8.00	&	CC3/AVTZ		&	-0.13	&	7.87	\\
	Benzene			&	1\,\ex{1}{A}{1g}{} $\ra$ 1\,\ex{1}{E}{2g}{}				&	exFCI/6-31+G(d)					&	8.40	&	CC3/AVTZ		&	-0.12	&	8.28	\\
					&	1\,\ex{1}{A}{1g}{} $\ra$ 2\,\ex{1}{A}{1g}{}				&	XMS-CASPT2/AVQZ					&	10.54	&	\cdash			&	\cdash	&	10.54	\\
	Beryllium		&	1\,\ex{1}{S}{}{} $\ra$ 1\,\ex{1}{D}{}{}					&	Ref.~\onlinecite{Galvez_2002}	&	7.06	&	\cdash			&	\cdash	&	7.06	\\
	Butadiene		&	1\,\ex{1}{A}{g}{} $\ra$ 2\,\ex{1}{A}{g}{}				&	exFCI/AVDZ						&	6.51	&	CC3/AVQZ		&	-0.01	&	6.50	\\
	Carbon dimer	&	1\,\ex{1}{\Sigma}{g}{+} $\ra$ 1\,\ex{1}{\Delta}{g}{}	&	exFCI/AVQZ						&	2.06	&	\cdash			&	\cdash	&	2.06	\\
					&	1\,\ex{1}{\Sigma}{g}{+} $\ra$ 2\,\ex{1}{\Sigma}{g}{+}	&	exFCI/AVQZ						&	2.40	&	\cdash			&	\cdash	&	2.40	\\
	Carbon trimer	&	1\,\ex{1}{\Sigma}{g}{+} $\ra$ 1\,\ex{1}{\Delta}{g}{}	&	exFCI/AVQZ						&	5.23	&	\cdash			&	\cdash	&	5.23	\\
					&	1\,\ex{1}{\Sigma}{g}{+} $\ra$ 2\,\ex{1}{\Sigma}{g}{+}	&	exFCI/AVQZ						&	5.86	&	\cdash			&	\cdash	&	5.86	\\
	Ethylene		&	1\,\ex{1}{A}{g}{} $\ra$ 2\,\ex{1}{A}{g}{}				&	exFCI/AVTZ						&	12.92	&	CC3/AVQZ		&	-0.36	&	12.56	\\
	Formaldehyde	&	1\,\ex{1}{A}{1}{} $\ra$ 3\,\ex{1}{A}{1}{}				&	exFCI/AVTZ						&	10.35	&	CC3/AVQZ		&	-0.01	&	10.34	\\
	Glyoxal			&	1\,\ex{1}{A}{g}{} $\ra$ 2\,\ex{1}{A}{g}{}				&	exFCI/AVDZ						&	5.48	&	CC3/AVQZ		&	+0.06	&	5.54	\\
	Hexatriene		&	1\,\ex{1}{A}{g}{} $\ra$ 2\,\ex{1}{A}{g}{}				&	CC3/AVDZ						&	5.77	&	PC-NEVPT2/AVQZ	&	-0.02	&	5.75	\\			
	Nitrosomethane	&	1\,\ex{1}{A}{}{'} $\ra$ 2\,\ex{1}{A}{}{'}				&	exFCI/AVTZ						&	4.76	&	CC3/AVQZ		&	-0.02	&	4.74	\\
	Nitroxyl		&	1\,\ex{1}{A}{}{'} $\ra$ 2\,\ex{1}{A}{}{'}				&	exFCI/AVQZ						&	4.32	&	\cdash			&	\cdash	&	4.32	\\
	Pyrazine		&	1\,\ex{1}{A}{g}{} $\ra$ 2\,\ex{1}{A}{g}{}				&	PC-NEVPT2/AVQZ					&	8.04	&	\cdash			&	\cdash	&	8.04	\\
					&	1\,\ex{1}{A}{g}{} $\ra$ 3\,\ex{1}{A}{g}{}				&	CC3/AVTZ						&	8.69	&	PC-NEVPT2/AVQZ	&	+0.00	&	8.69	\\	
	Tetrazine		&	1\,\ex{1}{A}{g}{} $\ra$ 2\,\ex{1}{A}{g}{}				&	PC-NEVPT2/AVQZ					&	4.60	&	\cdash			&	\cdash	&	4.60	\\	
					&	1\,\ex{1}{A}{g}{} $\ra$ 1\,\ex{1}{B}{3g}{}				&	PC-NEVPT2/AVQZ					&	6.14	&	\cdash			&	\cdash	&	6.14	\\	
					&	1\,\ex{1}{A}{g}{} $\ra$ 1\,\ex{3}{B}{3g}{}				&	PC-NEVPT2/AVQZ					&	5.51	&	\cdash			&	\cdash	&	5.51	\\	
	\end{tabular}
	\end{ruledtabular}
\end{table*}
\end{squeezetable}
%%% %%% %%%

%%%%%%%%%%%%%%%%%%%%%%%%
\subsection{
Theoretical best estimates
\label{sec:TBE}
}
%%%%%%%%%%%%%%%%%%%%%%%%
In Table \ref{tab:TBE}, we report TBEs for the vertical excitations considered in Table \ref{tab:2Ex}.
These TBEs are computed as $\Delta E_\text{R/SB} + \Delta E_\text{C/LB} - \Delta E_\text{C/SB}$, where $\Delta E_\text{R/SB}$ is the excitation energy computed with a reference (R) method in a small basis (SB), and $\Delta E_\text{C/SB}$ and $\Delta E_\text{C/LB}$ are excitation energies computed with a correction (C) method in the small and large basis (LB), respectively.
By default, we have taken as reference the exFCI excitation energies ($\Delta E_\text{R/SB}$) computed in the present study, while the basis set correction ($\Delta E_\text{C/LB} - \Delta E_\text{C/SB}$) is calculated at the CC3 level.
When the exFCI result is unavailable, we have selected, for each excitation separately, what we believe is the most reliable reference method. 
For most excitations (except the 1\,\ex{1}{A}{g}{} $\ra$ 2\,\ex{1}{A}{g}{} transition in ethylene), the basis set correction is small.
In the case of \ce{Be}, the value of Ref.~\onlinecite{Galvez_2002} is indisputably more accurate than ours.
For \ce{C2}, butadiene and hexatriene, we have not chosen the heat-bath CI results \cite{Holmes_2017, Chien_2018} as reference because these calculations were not performed at the same CC3 geometry.
However, these values are certainly outstanding references for their corresponding geometry.

%%% TABLE 3 %%%
\begin{squeezetable}
\begin{table}
	\caption{
	Mean absolute error (MAE), root mean square error (RMSE), as well as minimum (Min.) and maximum (Max.) absolute errors (with respect to exFCI) of CC3, CCSDT, CCSDTQ, CASPT2, PC-NEVPT2 and SC-NEVPT2 excitation energies.
	All quantities are given in eV. 	
	``Count'' refers to the number of transitions considered for each method.
	\label{tab:stat}
	}
	\begin{ruledtabular}
	\begin{tabular}{lddddd}
		Method	&	\mcc{Count}	&	\mcc{MAE}	&	\mcc{RMSE}	&	\mcc{Min.}	&	\mcc{Max.}	\\
		\hline
		\mc{6}{l}{All excitations}	\\
		CC3			&	39		&	0.78		&	0.90		&	0.00		&	1.46	\\
		CCSDT		&	37		&	0.40		&	0.46		&	0.00		&	0.74	\\
		CCSDTQ		&	19		&	0.03		&	0.05		&	0.00		&	0.12	\\
		CASPT2		&	39		&	0.03		&	0.11		&	0.01		&	0.27	\\
		PC-NEVPT2	&	39		&	0.07		&	0.18		&	0.00		&	0.97	\\
		SC-NEVPT2	&	39		&	0.07		&	0.12		&	0.01		&	0.34	\\
		\hline
		\mc{6}{l}{Excitations with $\%T_1 > 50\%$}	\\
		CC3			&	4		&	0.11		&	0.13		&	0.00		&	0.18	\\
		CCSDT		&	3		&	0.06		&	0.07		&	0.00		&	0.08	\\
		CCSDTQ		&	0		&	\cdash		&	\cdash		&	\cdash		&	\cdash	\\
		CASPT2		&	4		&	0.12		&	0.19		&	0.03		&	0.27	\\
		PC-NEVPT2	&	4		&	0.13		&	0.18		&	0.18		&	0.23	\\
		SC-NEVPT2	&	4		&	0.22		&	0.24		&	0.08		&	0.31	\\
		\hline
		\mc{6}{l}{Excitations with $\%T_1 < 50\%$}	\\
		CC3			&	35		&	0.86		&	0.95		&	0.00		&	1.46	\\
		CCSDT		&	34		&	0.42		&	0.48		&	0.00		&	0.74	\\
		CCSDTQ		&	19		&	0.03		&	0.05		&	0.00		&	0.12	\\
		CASPT2		&	35		&	0.02		&	0.10		&	0.01		&	0.25	\\
		PC-NEVPT2	&	35		&	0.07		&	0.18		&	0.00		&	0.97	\\
		SC-NEVPT2	&	35		&	0.06		&	0.10		&	0.01		&	0.34	\\
	\end{tabular}
	\end{ruledtabular}
\end{table}
\end{squeezetable}

%%% FIG 3 %%%
\begin{figure*}
	\includegraphics[height=0.33\linewidth]{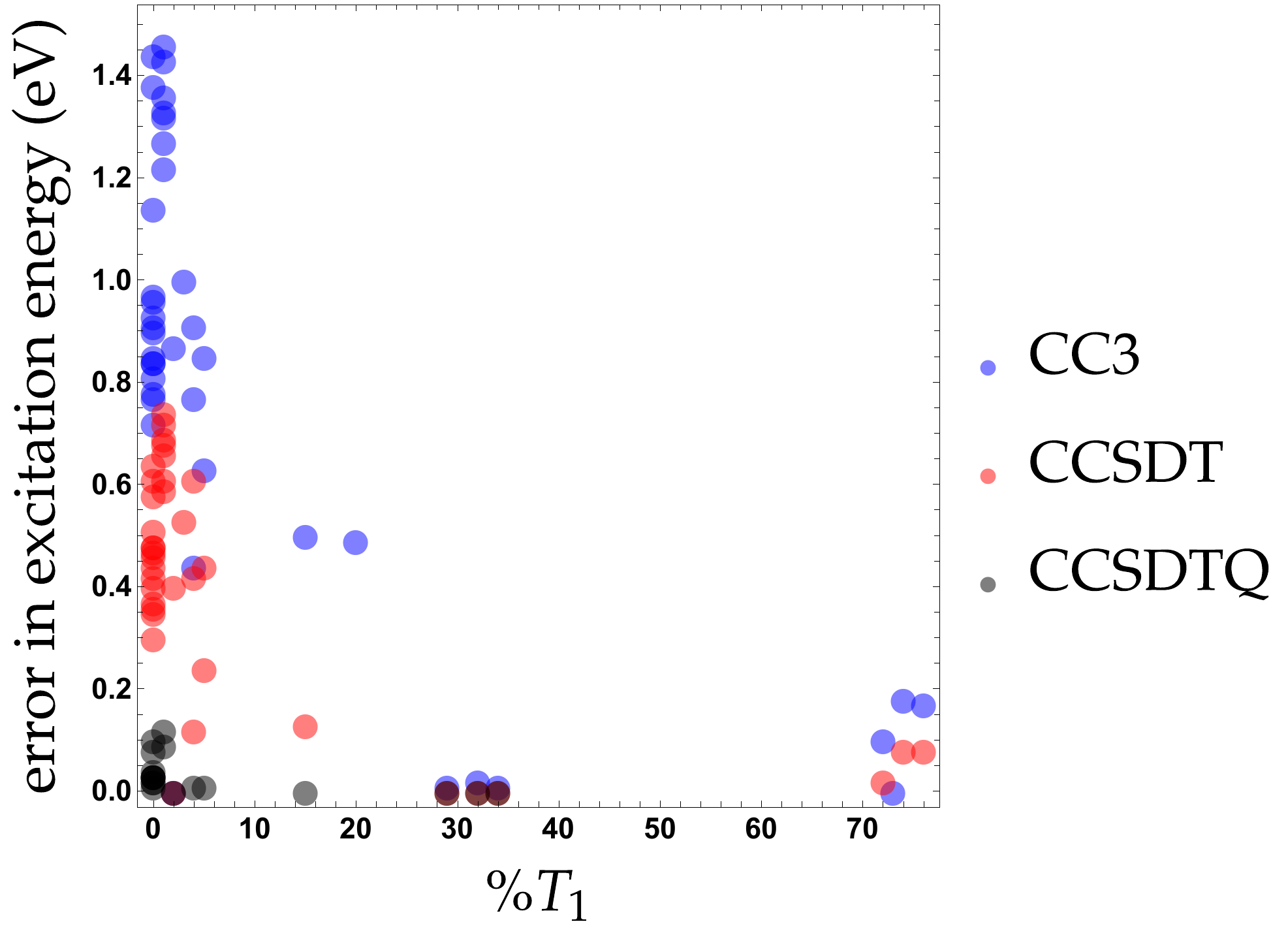}
	\includegraphics[height=0.33\linewidth]{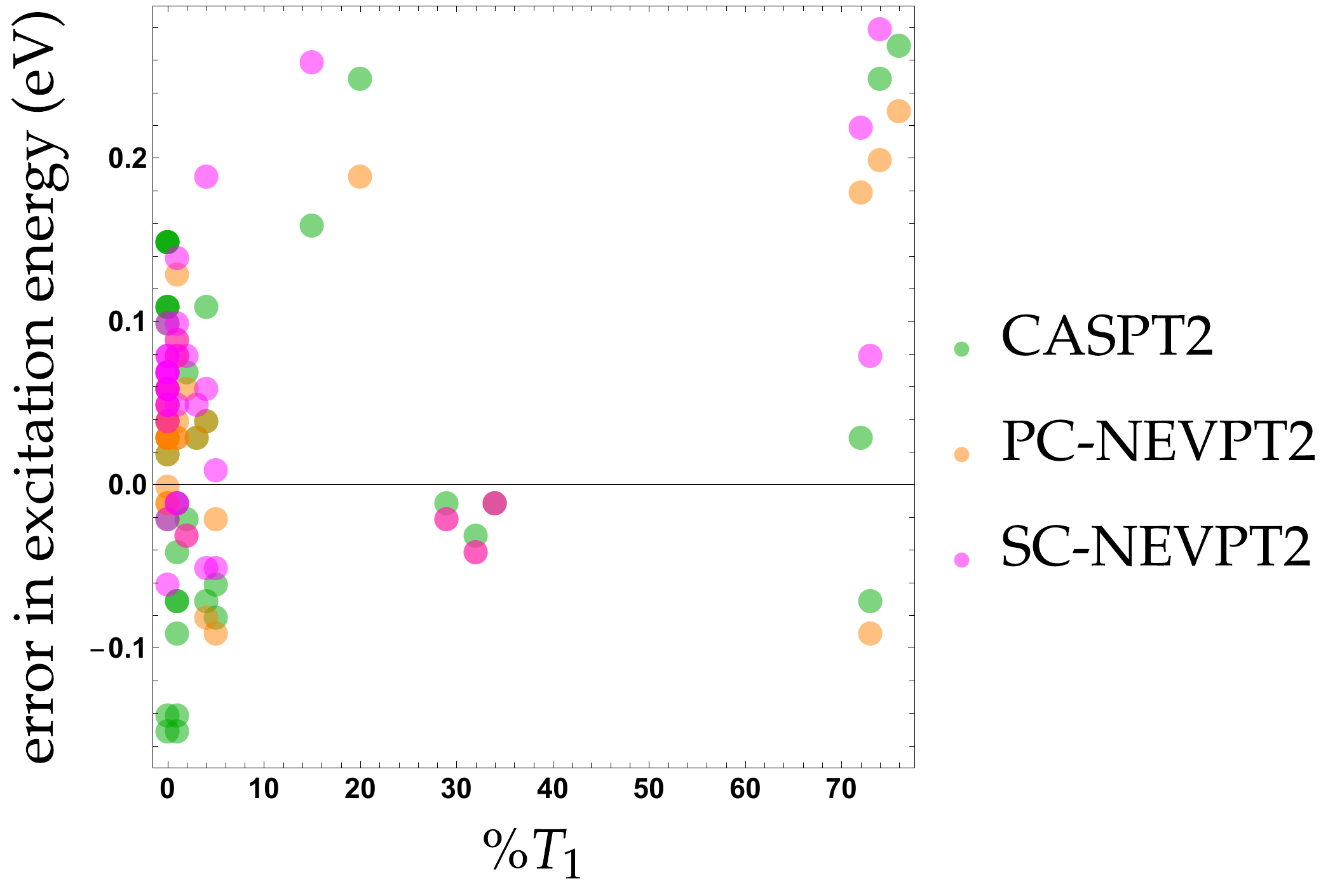}
	\caption{
	Error in excitation energies (in eV) with respect to exFCI as a function of the percentage of single excitation $\%T_1$ (computed at the CC3 level) for various molecules and basis sets.
	Left: CC3 (blue), CCSDT (red) and CCSDTQ (black).
	Right: CASPT2 (green), PC-NEVPT2 (orange) and SC-NEVPT2 (pink).
	Note the difference in scaling of the vertical axes.
	\label{fig:ExvsT1}
	}
\end{figure*}

%%%%%%%%%%%%%%%%%%%%%%%%
\section{
Conclusion
\label{sec:ccl}
}
%%%%%%%%%%%%%%%%%%%%%%%%

We have reported reference vertical excitation energies for 20 transitions with significant double excitation character in a set of 14 small- and medium-size compounds using a series of increasingly large diffuse-containing atomic basis sets (from Pople's 6-31+G(d) to Dunning's aug-cc-pVQZ basis).
Depending on the size of the molecule, selected configuration interaction (sCI) and/or multiconfigurational (CASSCF, CASPT2, (X)MS-CASPT2 and NEVPT2) calculations have been performed in order to obtain reliable estimates of the vertical transition energies.  

We have shown that the error obtained with CC methods including iterative triples can significantly vary with the exact nature of the transition. 
For ``pure'' double excitations (i.e.~for transition which do not mix with single excitations), the error in CC3 can easily reach \IneV{$1$} (and up to \IneV{$1.5$}), while it goes down to few tenths of an eV for more common transitions (like in butadiene, acrolein and benzene) involving a significant amount of singles.
This analysis is corroborated by Fig.~\ref{fig:ExvsT1} which reports the CC3, CCSDT, and CCSDTQ excitation energy errors with respect to exFCI as a function of the percentage of single excitation $\%T_1$ (computed at the CC3 level).  
A statistical analysis of these data is also provided in Table \ref{tab:stat} where one can find the mean absolute error (MAE), root mean square error (RMSE), as well as the minimum and maximum absolute errors associated with the CC3, CCSDT, and CCSDTQ excitation energies.  
For CC3, one can see a clear correlation between the magnitude of the error and the degree of double excitation of the corresponding transition.  
CC3 returns an overall MAE of \IneV{$0.78$} which drops to \IneV{$0.11$} when one considers solely excitations with $\%T_1 > 50\%$ (with a maximum error as small as \IneV{$0.18$}), but raises to \IneV{$0.86$} for excitations with $\%T_1 < 50\%$. 
Therefore, one can conclude that CC3 is a particularly accurate method for excitations dominated by single excitations which are ubiquitous, for instance, in compounds like butadiene, acrolein, hexatriene, and benzene derivatives.  
Indeed, according to our results, CC3 outperforms CASPT2 and NEVPT2 for these transitions (see below).
This corroborates the conclusions drawn in our previous investigation where we evidenced that CC3 delivers very small errors with respect to FCI estimates for small compounds. \cite{Loos_2018} 
A similar trend is observed with CCSDT at a lower scale: the overall MAE is \IneV{$0.40$} (a two-fold reduction compared to CC3), but $0.06$ and \IneV{$0.42$} for transitions with $\%T_1 > 50\%$ and $\%T_1 < 50\%$, respectively. 
As expected, more computationally demanding approaches like CCSDTQ (and beyond) yield highly accurate results even for genuine double excitations. 
For CCSDTQ, we have not been able to perform calculations on single-dominant excitations as such type of excitations does not seem to appear in small molecules.
From a general point of view, CC methods consistently overestimate excitation energies compared to exFCI.

The quality of the excitation energies obtained with multiconfigurational methods such as CASPT2, (X)MS-CASPT2, and NEVPT2 is harder to predict. 
We have found that the overall accuracy of these methods is highly dependent of the system and the selected active space. 
Note, however, that including the $\si$ and $\sis$ orbitals in the active space, even for transitions involving mostly $\pi$ and $\pis$ orbitals, can significantly improve the excitation energies.
The statistics associated with the CASPT2, PC-NEVPT2 and SC-NEVPT2 data are also provided in Table \ref{tab:stat} and depicted in Fig.~\ref{fig:ExvsT1}.
The overall MAE of CASPT2 is \IneV{$0.03$}, i.e., identical to CCSDTQ, while it is slightly larger for the two NEVPT2 variants (\IneV{$0.07$} for both of them).
However, their RMSE (which gives a bigger weight to large errors) is much larger.
Similar observations can be made for excitations with $\%T_1 < 50\%$, while for single-dominant excitations (i.e.~$\%T_1 > 50\%$), the MAEs in multiconfigurational methods are higher than in CC-based methods.
As a final comment, we note that the consistent overestimation of the exFCI excitation energies observed in CC methods does not apply to multiconfigurational methods.

We believe that the reference data reported in the present study will be particularly valuable for the future development of methods trying to accurately describe double excitations.

%%%%%%%%%%%%%%%%%%%%%%%%
\section*{Supporting Information}
%%%%%%%%%%%%%%%%%%%%%%%%
See {\SI} for geometries and additional information (including total energies) on the CC, multiconfigurational and sCI calculations.

%%%%%%%%%%%%%%%%%%%%%%%%
\begin{acknowledgements}
P.F.L.~would like to thank Emmanuel Giner and Jean-Paul Malrieu for NEVPT2-related discussions.
P.F.L.~and A.S.~would like to thank Nicolas Renon and Pierrette Barbaresco (CALMIP, Toulouse) for technical assistance.
D.J.~acknowledges the \emph{R\'egion des Pays de la Loire} for financial support. 
This work was performed using HPC resources from 
i) GENCI-TGCC (Grant No. 2018-A0040801738),
ii) CCIPL (\emph{Centre de Calcul Intensif des Pays de Loire}), 
iii) the Troy cluster installed in Nantes, and 
iv) CALMIP (Toulouse) under allocations 2018-0510, 2018-18005 and 2018-12158. 
\end{acknowledgements}
%%%%%%%%%%%%%%%%%%%%%%%%


\end{document}